\documentclass{tlp}

\usepackage{epsfig}
\usepackage{wrapfig}
\usepackage{subfigure}
\usepackage{rotating}
\usepackage{moreverb}
\usepackage{fancyvrb}
\usepackage{amsmath}
\usepackage{amssymb}
\usepackage[matrix,arrow,frame,curve]{xy}

\newcommand{\slgc}{\slg{\CT}}
\newcommand{\slg}[1]{SLG$^{#1}$}

\newcommand{\CT}{{\cal D}}
\newcommand{\Herbrand}{\mathcal{H}}
\newcommand{\chr}{\mathit{CHR}}
\newcommand{\project}[1]{\bar{\exists}_{#1}}
\newcommand{\ans}{\mathit{ans}}
\newcommand{\vars}[1]{\ensuremath{\mathit{vars}(#1)}}
\newcommand{\true}{\mathit{true}}
\newcommand{\false}{\mathit{false}}
\newcommand{\Prolog}{\ensuremath{{\cal H}}}
\newcommand{\Prog}{{\ensuremath{\mathcal{P}}}}

\newcommand{\num}[2]{#1\##2}
\newcommand{\BT}{\ensuremath{\mathcal{D}_b}}
\newcommand{\plusplus}{\mathrel+\joinrel\mathrel+}

\newcommand{\numchr}[1]{\ensuremath{\mathit{chr}(#1)}}
\newcommand{\numid}[1]{\ensuremath{\mathit{id}(#1)}}

\newcommand{\transfo}{\mathbf{T}}

\newcounter{example}
\renewcommand{\theexample}{\arabic{example}}

\numberwithin{equation}{section}

\newenvironment{example}[0]{\ \newline \refstepcounter{example}\noindent\textbf{Example \theexample\ }}{$\Box$\newline}

\newtheorem{definition}{Definition}

\begin{document}

\title{TCHR: a framework for tabled CLP}
\author[T. Schrijvers et al.]{TOM SCHRIJVERS\thanks{Research Assistant of the fund for Scientific Research - Flanders (Belgium)(F.W.O. - Vlaanderen)} and BART DEMOEN \\
       Dept. of Computer Science, K.U.Leuven, Belgium \\
       \email{\{toms,bmd\}@cs.kuleuven.ac.be}
       \and
       DAVID S. WARREN\\
        Dept. of Computer Science, State University of New York at Stony Brook, USA \\
       \email{warren@cs.sunysb.edu}
}
 \submitted{25 September 2006 }
 \revised{19 July 2007}
 \accepted{19 Dec 2007}
\maketitle

\begin{abstract}
Tabled Constraint Logic Programming is a powerful execution mechanism
for dealing with Constraint Logic Programming without worrying about
fixpoint computation. Various applications, e.g in the fields of
program analysis and model checking, have been proposed. Unfortunately,
a high-level system for developing new applications is lacking, and 
programmers are forced to resort to complicated ad hoc solutions.

This papers presents TCHR, a high-level framework for tabled Constraint
Logic Programming. It integrates in a light-weight manner Constraint
Handling Rules (CHR), a high-level language for constraint solvers, with
tabled Logic Programming. The framework is easily instantiated with new
application-specific constraint domains. Various high-level operations can
be instantiated to control performance. In particular, we propose a novel,
generalized technique for compacting answer sets.
%
\end{abstract}



\section{Introduction}

The notion of tabled Constraint Logic Programming (CLP) originates from the
{\em constraint databases} community \cite{kanellakis:cql}. In an ordinary
database, data is stored in relations of atomic values. Constraint databases
generalize atomic values to constraint variables: a field is restricted
to a range of values rather than a single value. This allows for more
compact representations than explicitly enumerating the atomic values. 
DATALOG, a formalism for reasoning about ordinary
databases and queries in particular, is generalized to DATALOG$^\CT$ for
this purpose. Just as DATALOG is a restricted form of Logic Programming,
DATALOG$^\CT$ is a restricted form of Constraint Logic Programming. The
restrictions enforce programs to have finite interpretations. 

Due to the finiteness properties, queries on DATALOG$^\CT$ programs
can be resolved by bottom-up computation rather than the usual top-down
goal-directed computation of CLP. The former has the advantage that it
terminates for DATALOG$^\CT$ programs, whereas the latter may get stuck
in infinite loops. However, a goal-directed approach usually obtains the
desired result much faster and uses less space. For this reason, Toman
\cite{toman:memo} proposed a compromise: tabling. Tabling is an LP
technique for improving the termination properties of the goal-directed
approach through the memoization of intermediate results. By generalizing
tabled DATALOG$^\CT$ to Tabled CLP, we benefit from both the generalized
expressivity of CLP and the improved termination properties of tabling.

A number of different applications have been proposed for tabled
DATALOG$^\CT$ and Tabled CLP. Toman himself considers it an alternative
approach to implementing
abstract interpretation \cite{toman:analysis}: constraints abstract
concrete values and tabling takes care of fixpoints. Various applications
in the context of model checking have been developed \cite{mcrtpre,mcrt,giri}:
constraints impose restrictions on parameters in parametrized models while
tabling takes care of cycles in the model graphs.


The above establishes a clear need for Tabled CLP, but let us consider the
availability of Tabled CLP systems. It turns out that a user-friendly and
comprehensive system for developing new Tabled CLP applications is missing
completely.
The above-mentioned model checking applications (e.g. \cite{mcrtpre,mcrt,giri}) have
adapted an existing tabled logic programming system, XSB \cite{xsb}, with constraint
programming facilities in various ad hoc and laborious ways.

At first, XSB developers resorted to interfacing with foreign language libraries or
implementing constraint solvers in XSB itself with a close coupling of
constraint solver and application as a consequence. For instance, the initial
feasibility study of a real-time model checking system used a meta interpreter
written in XSB to deal with constraints (see \cite{mcrtpre}). The subsequent
full system implements an interface between XSB and the POLINE polyhedra-based
constraint solver library, and passes around handles to the constraint store
in the XSB program (see \cite{mcrt}). At a later stage this real time model
checking application used distance bound matrices implemented in XSB itself
(see \cite{giri}).

In an attempt to facilitate the use of constraints, XSB was extended with
attributed variables \cite{bao:tclp}. Attributed variables
\cite{christian:attvars} are Prolog language feature widely used for
implementing constraint solvers. It allows to associate data with unbound
variables, manipulate it at will and also to interrupt the unification of these
variables. Unfortunately, constraint solvers are complex programs and even
with attributed variables it can be a daunting task to implement them. 

In order to substantially lower the threshold for tabled CLP, a high-level
formalism is needed for writing new constraint solvers and integrating them in
a tabled logic programming system. In this work we present such a formalism:
Tabled Constraint Handling Rules, or TCHR for short.

TCHR is a high-level framework for developing new constraint solvers in a
tabled logic programming environment. It integrates Constraint Handling Rules
(CHR) \cite{thom:main}, an established high-level formalism for writing new
constraint solvers, and tabled logic
programming. The framework offers a number of default operations that can
be specialized by instantiations to control both semantics and performance.

A practical implementation of the framework is presented: the integration  of
K.U.Leuven CHR in XSB. The integration shows how a tabled constraint logic
programming system can be obtained from a Constraint Logic Programming and a
tabled logic programming system with little impact on either. Although we have
chosen XSB as our particular tabled logic programming system, we believe that
our ideas readily apply to other table-based LP systems. 


In summary, the major contributions of this work are:
\begin{itemize}
\item a high-level framework for developing new constraint solvers in a
      tabled logic programming system, 
\item a practical implementation of the framework in terms of K.U.Leuven
      CHR and XSB, and
\item a novel, generalized approach for answer set reduction. 
\end{itemize}

The CHR-XSB integration, we believe, combines both the bottom-up and top-down fixpoint
computations, the superior termination properties of XSB and the constraint
programming capabilities of CHR. This combined power enables programmers to
easily write highly declarative programs that are easy to maintain and extend.


\textbf{Overview}\hspace{3mm}
The rest of this text is structured as follows. First, in Sections
\ref{sec:slg_ct} and \ref{sec:chr} we provide basic technical background
knowledge on tabled execution of Constraint Logic Programs and Constraint
Handling Rules, respectively. 

Section \ref{sec:tabling} outlines our contribution: a framework for
tabled CLP system integrated in terms of SLG and Constraint Handling Rules. 
Subsequent sections discuss in more detail the different options and
operations of the framework: call abstraction (Section~\ref{sec:abstraction}),
answer projection (Section~\ref{sec:projection}) and answer set optimization
(Section~\ref{sec:asopt}).

Finally, Section~\ref{sec:related} discusses related and possible future work,
and Section~\ref{sec:conclusion} concludes.

But first we end this introduction with a small motivating example
from the domain of model checking:
\begin{example}
{\em Data-independent systems} \cite{wolper86} manipulate data variables over unbounded domains
but have a finite number of control locations. Such systems can be modeled
as {\em extended finite automata} \cite{beata:efa}: finite automata with {\em guards} on the transitions
and {\em variable mapping} relations between source and destination locations. They are useful
for modelling and subsequently checking e.g. buffers and protocols.

A simple example of such a data-independent system, modeled in CLP(FD), is:
\begin{Verbatim}
        edge(a,b,Xa,Xb) :- Xa < 10, Xb = Xa.
        edge(b,a,Xb,Xa) :- Xb > 0, Xa = Xb + 1.
        edge(b,c,Xb,Xc) :- Xb > 3, Xc = Xb.
\end{Verbatim}
This system has three control locations \texttt{a,b,c} each with one variable,
respectively \texttt{Xa,Xb,Xc}. Each \texttt{edge/4} clause represents an edge
in the system: a possible transition from one control location (the source) to
another (the destination). The inequality constraint in each clause {\em
guards} the transition, and the equality constraint relates the source variable
to the destination variable (the variable mapping).

Suppose that we are interested in whether location \texttt{c} is reachable from
location \texttt{a}, and for which values of the parameter \texttt{Xa}.
Let us define a reachability predicate:
\begin{Verbatim}
        reach(A,A,X).
        reach(A,C,X) :-
          edge(A,B,X,NX),
          reach(B,C,NX).     
\end{Verbatim}
Then our reachability question is captured by the query \texttt{?-
reach(a,c,X)}. In order to answer this query, tabling is required to avoid
the non-termination trap of the \texttt{a-b} cycle in the graph. At the same time,
constraints allow a compact symbolic representation of the  
infinite search space for \texttt{X}.
Without a good interaction between both tabling and constraints, we would not
be able to obtain as concise a solution
as {\tt 0 < X < 10} with so little effort. 
\end{example}

%
%

\section{Tabled Constraint Logic Programs}\label{sec:slg_ct}

In this section we cover the basics of tabled Constraint Logic Programming.
First, the syntax of Constraint Logic Programs is presented in Section
\ref{sec:slg_ct:syntax}. Next, Section \ref{sec:slg_ct:domains} explains
about the constraints part of CLP: the constraint domain. Finally, Section
\ref{sec:slg_ct:semantics} presents the (operational) semantics \slgc\
of Tabled Constraint Logic Programs.

\subsection{Syntax of Constraint Logic Programs}\label{sec:slg_ct:syntax}

A {\em Constraint Logic Program} consists of a number of rules, called {\em clauses},
of the form:
        \[ H \texttt{:-} C, L_1\texttt{,}\ldots\texttt{,}L_n\texttt{.} \] 
where $n \geq 0$ and $H$ is an atom, $C$ is a constraint and $L_1,\ldots,L_n$ are {\em literals}. A
literal is either an atom $A$ or a negated atom $\neg A$.

$H$ is called the {\em head} of the clause and $C, L_1,\ldots,L_n$ is called
the {\em body}. The comma ``\texttt{,}'' is called {\em conjunction} as it
corresponds with logical conjunction in the semantics of Constraint Logic Programs.

The atoms are constructed from predicate symbols $p/n$ and variables. Their
meaning is defined by the Constraint Logic Programming itself.
The syntax and semantics of constraints
is defined by the constraint domain $\CT$ (see Section \ref{sec:slg_ct:domains}).


If all the literals in the body are positive, the clause is a {\em definite
clause}. A {\em normal clause} is a clause that may also contain negative
literals. A {\em definite Constraint Logic Program} consists of definite clauses only,
while a {\em normal Constraint Logic Program} has normal clauses. 
From now one we will only consider definite programs and address them as
programs for short.

\subsection{Constraint Domains}\label{sec:slg_ct:domains}

A constraint solver is a (partial) executable implementation of a
constraint domain. A {\em constraint domain} $\CT$ consists of a set 
$\Pi$ of constraint symbols, a logical theory $\mathcal{T}$ and for
every constraint symbol $c/n \in \Pi$ a tuple of value sets $\langle
V_1,\ldots,V_n \rangle$.  A {\em primitive constraint} is constructed from a
constraint symbol $c/n$ and for every argument position $i$ ($1 \leq i \leq n$)
either a variable or a value from the corresponding value set $V_i$, similar to
the way an atom is constructed in a logic program.

A {\em constraint} is of the form $c_1 \wedge \ldots \wedge c_n$ where $n \geq
0$ and $c_1, \ldots, c_n$ are primitive constraints. Two distinct constraints
are $\true$ and $\false$. The former always holds and the latter never
holds. The empty conjunction of constraints is written as $\true$.

The logical theory $\mathcal{T}$ determines what constraints hold and what
constraints do not hold. Typically, we use $\CT$ to also refer specifically to
$\mathcal{T}$. For example $\CT \models c$ means that under the logical theory
$\mathcal{T}$ of constraint domain $\CT$ the constraint $c$ holds.

A {\em valuation} $\theta$ for a constraint $C$ is a variable substitution that
maps the variables in $\vars{C}$ onto values of the constraint domain $\CT$.
If $\theta$ is a valuation for $C$, then it is a {\em solution} for $C$ if
$C\theta$ holds in the constraint domain $\CT$, i.e. $\CT \models C\theta$.
A constraint $C$ is {\em satisfiable} if it has a {\em solution}; otherwise it
is {\em unsatisfiable}. Two constraints $C_1$ and $C_2$ are {\em equivalent},
denoted $\CT \models C_1 \leftrightarrow C_2$, if and only if they have the
same solutions.

A constraint domain of particular interest is the Herbrand domain $\Herbrand$.
Its only constraint symbol is term equality $=/2$, which ranges over Herbrand
terms. Plain Logic Programming can be seen as a specialized form of Constraint
Logic Programming over the Herbrand domain.

Two problems associated with a constraint $C$ are the {\em solution problem},
i.e. determining a particular solution, and the {\em satisfaction problem},
i.e. determining whether there exists at least one solution.  An algorithm for
determining the satisfiability of a constraint is called a {\em constraint
solver}. Often a solution is produced as a by-product. A general technique
used by many constraint solvers is to repeatedly rewrite a constraint into an
equivalent constraint until a {\em solved form} is obtained. A constraint in
solved form has the property that it is clear whether it is satisfiable or not.
See \cite{stuckey:clp} for a more extensive introduction to constraint
solvers. 
\subsection{Semantics of Constraint Logic Programs}\label{sec:slg_ct:semantics}

In a survey of Constraint Logic Programming (CLP) \cite{jaffar:clp} various
forms of semantics are listed for Constraint Logic Programs: logic semantics based on
Clark completion\cite{clark:completion}, fixpoint semantics \cite{jaffar:fixpoint} as well
as a new framework for top-down and bottom-up operational semantics. 


The CLP fixpoint semantics are defined, in the usual way, as the fixpoint
of an extended immediate consequence operator.
\begin{definition}[CLP Immediate Consequence Operator]
The one-step consequence function $T_{P}^{\CT}$ for a CLP program $P$
with constraint domain $\CT$ is defined as:
\[ \begin{array}{ll}
T_P^{\CT}(I) = \{ p(\bar{d}) | & 
	p(\bar{x}) \leftarrow c,b_1,\ldots,b_n \in P, \\
&	\exists v . v \text{ is a valuation on } \CT: \\
&       \hspace{1cm} \CT \models v(c), \\
&       \hspace{1cm} v(\bar{x}) = \bar{d}, \\
& 	\hspace{1cm} \forall i: 1 \leq i \leq n \Rightarrow v(b_i) \in I \}
\end{array}\]
\end{definition}

A goal-directed execution strategy, \slgc, using tabling for the above
CLP fixpoint semantics has been developed by Toman in \cite{toman:memo}.
This \slgc\ semantics encompasses the best of both top-down and bottom-up
operational semantics: it is {\em goal-directed} like top-down evaluation
and has the favorable {\em termination} properties like bottom-up evaluation.

\subsubsection{Basic \slgc\ Semantics}

The \slgc\ semantics makes two assumptions about the constraint domain $\CT$.
Firstly, $\CT$ includes
a projection operation that returns a disjunction of constraints:
$\project{T}C = \bigvee_i C_i$. The notation
$C_j \in \project{T}C$ is used to state that $C_j$ is one of the disjuncts
in this disjunction. Secondly, it is assumed that a relation $\leq_\CT$
is provided. This relation should be at least as strong as implication, i.e.
\[ \forall C_1,C_2: C_1 \leq_\CT C_2 ~~\Rightarrow~~ \CT \models C_1 \rightarrow C_2 \]

\slgc\ is formulated in terms of four resolution  (or rewriting) rules, listed
in Table \ref{tab:tabling:slgc}. These rules either expand existing tree
nodes or create new root nodes. There are four different kinds of tree nodes:
$\texttt{root}(G;C)$, $\texttt{body}(G;B_1,\ldots,B_k;C)$,
$\texttt{goal}(G;B,C';B_2\ldots,B_k;C)$ and $\texttt{ans}(G;A)$
where $G$ is an atom, $B1,\ldots,B_k$ are literals, and $C,C',A$ are constraints\footnote{Also known as {\em constraint stores}.} in $\CT$.

An \slgc\ tree is built from a $\texttt{root}(G;C)$ node using the resolution
rules.
An \slgc\ forest is a set of \slgc\ trees.
The meaning of the different resolution rules is the following: 
\begin{itemize}
\item The {\em Clause Resolution} rule expands a root node: for every
      matching clause head a body node is created containing the clause's body literals. 
\item If there is at least one literal in a body node, it is expanded by
      the {\em Query Projection} rule into goal nodes.  This rule selects a
      literal to be resolved. The given {\em Query Projection} rule implements
      a left-to-right selection strategy, which is common to most LP systems,
      including XSB. However, any other strategy is valid as well. The
      current constraint store $C$ is projected onto the selected literal's
      variables, yielding only the constraints relevant for that literal. As
      the projection yields a disjunction of constraints, one goal node is
      created for every disjunct.
\item If there is no literal in a body node, it is expanded by the {\em Answer Projection}
      rule into a number of answer nodes. For this purpose the current constraint store
      $C$ is projected onto the goal's variables, retaining only those constraints relevant
      to the goal. In this way variables local to the chosen clause's body are eliminated.
\item A goal node is expanded into new body nodes by the {\em Answer
      Propagation} rule.
      This rule substitutes the selected literal by its answers: the
      selected literal's answer constraint stores are incorporated in the
      current store.
\end{itemize}
      
Note that the {\em Answer Propagation} and {\em Answer Projection}
rules cooperate: whenever a new answer is produced, it is propagated to all
the nodes that have already been resolved using answers from this tree.
Also the {\em Answer Propagation} rule is responsible for creating new \slgc\ trees:
when no tree with a root node that subsumes the goal ($B,C'$) to be resolved can be found,
a node $\texttt{root}(B,C')$ is created to start a separate tree.

Finally, a query in the SLG$^\CT$ formalism is a tuple $(G,C,P)$ where
$\vars{C} \subseteq \vars{G}$ and all arguments of $G$ are variables.
The \slgc\ resolution rules are used for query evaluation as follows:
\begin{enumerate}
\item create an \slgc\ forest containing a single tree $\{\texttt{root}(G,C)\}$,
\item expand the leftmost node using the resolution rules as long as they can be applied, and
\item return the set $\ans(G,C)$ as the answers for the query.
\end{enumerate}
\begin{definition}(Answer Set)
The answer set $\ans(G,C)$ is the set of all $A$ such that $\texttt{ans}(G;A) \in \mathit{slg}(G,C)$,
where $\mathit{slg}(G,C)$ is the \slgc\ tree rooted at $\texttt{root}(G,C)$.
\end{definition}


\begin{table}
{\small
\renewcommand{\arraystretch}{1.3}
\begin{tabular}{|@{\hspace{1mm}}r@{}l@{\hspace{-5pt}}l@{}|}
\cline{1-3}
\multicolumn{1}{|c}{\normalsize Parent} & \multicolumn{1}{c}{\normalsize Children} & \multicolumn{1}{c|}{\normalsize Conditions} \\ 

\cline{1-3}
\rule{0pt}{15pt}\framebox{Clause Resolution} & & \\
$\texttt{root}(G;C)$ &

$
\left\{\begin{array}{c}
\texttt{body}(G;B_1^1,\ldots,B_{k_1}^1;C \wedge \theta \wedge D^1) \\
\vdots \\
\texttt{body}(G;B_1^l,\ldots,B_{k_l}^l;C \wedge \theta \wedge D^l) \\
\end{array}\right.
$ 

& 
 
\begin{minipage}{0.3\textwidth}
for all $0 < i \leq l$ such that \\
$G' \rightarrow D^i, B_1^i,\ldots,B_{k_i}^i$ \\
and $\theta \equiv (G = G')$ \\
and $C \wedge \theta \wedge D^i$ is satisfiable \\
\end{minipage}
\\

\cline{1-3}
\rule{0pt}{15pt}\framebox{Query Projection} & & \\
$\texttt{body}(G;B_1,\ldots,B_k;C)$ &

$
\left\{\begin{array}{c}
\texttt{goal}(G;B_1,C_1;B_2\ldots,B_k;C) \\
\vdots \\
\texttt{goal}(G;B_1,C_l;B_2\ldots,B_k;C) \\
\end{array}\right.
$ 

& 
 
\begin{minipage}{0.3\textwidth}
for all $C_i \in \project{B_1}C$
\end{minipage}
\\
\cline{1-3}
\rule{0pt}{15pt}\framebox{Answer Propagation} & & \\
$\texttt{goal}(G;B_1,C_1;B_2\ldots,B_k;C)$ &

$
\left\{\begin{array}{c}
\texttt{body}(G;B_2,\ldots,B_k;C \wedge \theta \wedge A_1) \\
\vdots \\
\texttt{body}(G;B_2,\ldots,B_k;C \wedge \theta \wedge A_l) \\
\end{array}\right.
$ 

& 
 
\begin{minipage}{0.3\textwidth}
for all $A_i \in \ans(B',C')$ \\
where $\theta \equiv (B' = B_1)$ \\
and $C_1 \wedge \theta \leq_{\CT} C'$ \\
and $C \wedge \theta \wedge A_i$ is satisfiable
\end{minipage}
\\
\cline{1-3}
\rule{0pt}{15pt}\framebox{Answer Projection} & & \\
$\texttt{body}(G;\Box;C)$ &

$
\left\{\begin{array}{c}
\texttt{ans}(G;A_1) \\
\vdots \\
\texttt{ans}(G;A_l) \\
\end{array}\right.
$ 

& 
 
\begin{minipage}{0.3\textwidth}
for all $A_i \in \project{G}C$ 
\end{minipage}
\\
\cline{1-3}
\end{tabular}
}
\caption{SLG$^\CT$ resolution rules}\label{tab:tabling:slgc}
\end{table}
\subsubsection{An Example} 

Let us consider the following very simple CLP program $P$:
\begin{Verbatim}
        p(X) :- X = 1-Y, q(Y).
        p(X) :- X = 0.

        q(X) :- true, p(X).
\end{Verbatim}
The constraint domain is that of domain integers. The supported basic
constraint \texttt{=/2} is equality of arithmetic expressions.

Figure \ref{fig:slgc:ex} depicts the \slgc\ forest for the query $(p(U);\true;P)$.
The full arrows represent the \slgc\ tree branches, whereas the dashed arrow indicates
the start of a new tree and the dotted arrows indicate the propagation of new answers.
Each arrow is labeled with its step number.

The answer set $\ans(p(U),\true)$ consists of two answers: $U = 0$ and $U = 1$.

\begin{figure}
\begin{center}
\framebox{
$\xymatrixcolsep{0.0cm}
\xymatrix{
& \texttt{root}(p(U);\true)\ar[dl]_{(1:CR)}\ar[dr]^{(1:CR)} & \\
\texttt{body}(p(U);q(Y);U = 1-Y)\ar[d]_{(2:QP)} & & \texttt{body}(p(U);\Box;U = 0)\ar[d]^{(6:APj)} \\
\texttt{goal}(p(U);q(Y),\true;\Box;U = 1-Y) \ar[d]^{(9:APp)} \ar@{-->}[dddr]^{(3)}  \ar[dr]^{(13:APp)} & & \texttt{ans}(p(U);U=0) \ar@/^3pc/@{.>}[ddddddll]^<<<{(7)}\\
\texttt{body}(p(U);\Box;U = 1-Y \wedge Y=0) \ar[d]^{(10:Apj)} & \texttt{body}(p(U);\Box;U = 1-Y \wedge Y=1) \ar[d]^{(14:APj)} \\
\texttt{ans}(p(U);U=1) \ar@/_4pc/@{.>}[ddddrr]^<<<{(11)}                          & \texttt{ans}(p(U);U=0)\\
& \texttt{root}(q(Y);\true)      \ar[d]^{(4:CR)} & \\
& \texttt{body}(q(U);p(U);\true) \ar[d]^{(5:QP)} & \\
& \texttt{goal}(q(U);p(Y),\true;\Box;\true) \ar[dl]^{(7:APp)} \ar[dr]^{(11:APp)} \\
\texttt{body}(q(U);\Box;U=0) \ar[d]^{(8:APj)} & & \texttt{body}(q(U);\Box;U=1) \ar[d]^{(12:APj)} \\
\texttt{ans}(q(U);U=0) \ar@/^4pc/@{.>}[uuuuuu]^{(9)} & & \texttt{ans}(q(U);U=1) \ar@/_4pc/@{.>}[uuuuuul]_{(13)} \\
}
$
}
\end{center}
\caption{Example \slgc\ forest}\label{fig:slgc:ex}
\end{figure}
\subsubsection{\slgc\ Optimizations}

Several optimizations to the rewriting formulas have been proposed by Toman, of
which one, {\em Query Projection}, is of particular interest to us. The
optimization allows for more general goals than strictly necessary to be
resolved. In this way fewer goals have to be resolved, as distinct specific
queries can be covered by the same general goal.

Table \ref{tab:tabling:slgc:opt} lists the modified
{\em Query Projection} rule, called {\em Optimized Query Projection}.
\begin{table}
{\small
\renewcommand{\arraystretch}{1.3}
\begin{tabular}{|r@{}ll|}
\cline{1-3}
\multicolumn{1}{|c}{\normalsize Parent} & \multicolumn{1}{c}{\normalsize Children} & \multicolumn{1}{c|}{\normalsize Conditions} \\ 

\cline{1-3}
\multicolumn{2}{|l}{\rule{0pt}{15pt}\framebox{Optimized Query Projection}} & \\
$\texttt{body}(G;B_1,\ldots,B_k;C)$ &

$
\left\{\begin{array}{c}
\texttt{goal}(G;B_1,C_1;B_2\ldots,B_k;C) \\
\vdots \\
\texttt{goal}(G;B_1,C_l;B_2\ldots,B_k;C) \\
\end{array}\right.
$ 

& 
 
\begin{minipage}{0.3\textwidth}
$\CT \models \project{B_1}C \rightarrow C_1 \vee \ldots \vee C_l$ \\
for some $C_1,\ldots,C_l$
\end{minipage}
\\
\cline{1-3}
\end{tabular}
}
\caption{Optimized {\em Query Projection} for SLG$^\CT$ resolution}\label{tab:tabling:slgc:opt}
\end{table}

A second important optimization is a modified version of the answer set definition:
\begin{definition}(Optimized Answer Set)\label{def:tabling:slgc:entailment}
The optimized answer set of the query $(G,C,P)$, denoted $\ans(G,C)$, is the set of all $A$ such that $\texttt{ans}(G;A)
\in \mathit{slg}(G,C)$ and no $A'$ is already in $\ans(G,C)$
for which $A \leq_{\CT} A'$.
\end{definition}
This alternative definition allows for answers to be omitted if they are already
entailed by earlier more general answers. While logically the same answers
are entailed, the set of answers is smaller with the new definition.

Note that SLG, the operational semantics of tabled Logic Programming, is
in fact a specialized form of the \slgc\ semantics for the Herbrand domain.
Several implementations of SLG exist, including XSB. The topic of this paper,
the integration of CHR with tabled execution, is in effect an implementation
of \slgc\ for arbitrary $\CT$ defined by a CHR program.

In \cite{toman:wfs} Toman has also extended his work to a goal-directed
execution strategy for CLP programs with negation. This extension realizes
the well-founded semantics. An implementation of this extension is not
covered by our work. It imposes additional requirements on the constraint
solver: a finite representation of the negation of any constraint should exist.
Moreover, the detection of loops through negation requires a more complicated
tabling mechanism.

\section{Constraint Handling Rules}\label{sec:chr}

In this section we give a brief overview of Constraint Handling Rules
(CHR)~\cite{thom:main,thom:book}.

\subsection{Syntax of CHR}

We use two disjoint sets of predicate symbols for two different
kinds of constraints: {built-in (pre-defined) constraint symbols} 
which are solved by a given constraint solver,
and
{CHR (user-defined) constraint symbols} 
which are defined by the rules in a CHR program.
%
%
%
There are three kinds of rules:

\begin{center}
\begin{tabular}{ll}
{\em Simplification rule:} 	& ${\mathit Name} \texttt{ @ } H 		\texttt{ <=> } C \texttt{ | } B,$ \\
{\em Propagation rule:} 	& ${\mathit Name} \texttt{ @ } H 		\texttt{ ==> } C \texttt{ | } B,$ \\
{\em Simpagation rule:} 	& ${\mathit Name} \texttt{ @ } H \texttt{ \textbackslash\ } H' 	\texttt{ <=> } C \texttt{ | } B,$ \\
\end{tabular}
\end{center}

\noindent where {\em Name} is an optional, unique identifier of a rule, the {\em head}
$H$, $H'$ is a non-empty comma-separated conjunction of CHR constraints, the
{\em guard} $C$ is a conjunction of built-in constraints, and the {\em body} $B$
is a goal. A {\em query} is a conjunction of built-in and CHR
constraints.
%
%
A trivial guard expression ``\texttt{true |}'' can be omitted from a rule.
%
%
The head of a simplification rule is called a {\em removed} head, as the rule
replaces its head by its body. Similarly, the head of a propagation rule is
called a {\em kept} head, as the rule adds its body in the presence of its head.
Simpagation rules abbreviate simplification rules of the form
${\mathit Name} \texttt{ @ } H, H' \texttt{ <=> } C \texttt{ | } H, B$, i.e.
$H$ is a kept head and $H'$ a removed head.
A CHR program $\Prog$ consists of an ordered set of CHR rules.

%

%
%
%

\subsection{Operational Semantics of CHR}\label{sec:chr:semantics}

The formal operational semantics of CHR is given in terms of a state transition system
in Figure \ref{fig:chr:semantics}. The program state is an indexed 4-tuple $ \langle G, S, B, T \rangle_n$.
The first part of tuple, the \emph{goal} $G$ is the multiset of constraints to be rewritten to solved form.
The CHR constraint \emph{store} $S$ is the multiset of \emph{identified} CHR  
constraints that can be matched with rules in the program $P$.
An \emph{identified} CHR constraint\index{CHR constraint!identified} $\num{c}{i}$ is a CHR constraint 
$c$ associated with some unique integer $i$, the {\em constraint identifier}.
This number serves to
differentiate among copies of the same constraint.
We introduce the functions $\numchr{\num{c}{i}} = c$ and $\numid{\num{c}{i}} = i$,
and extend them to sequences, sets and multisets of identified CHR constraints in
the obvious manner, e.g. $\numchr{S} = \{ c ~|~ \num{c}{i} \in S\}$.

The \emph{built-in constraint store} $B$
is the conjunction of all built-in constraints that have been passed to the
underlying solver.
Since we will usually have no information about the internal representation of 
$B$,
we will model it as an abstract logical conjunction of constraints.
The \emph{propagation history} $T$ is a set of sequences, 
each recording the identities of the CHR constraints that fired a rule,
and the name of the rule itself. 
This is necessary to prevent trivial non-termination for propagation rules:
a propagation rule is allowed to fire on a set of constraints only if
the constraints have not been used to fire the same rule before.
Finally, the counter $n$ represents the next free integer that can be 
used to number a CHR constraint.

Given an initial query $G$, the \emph{initial program state} is:
$
        \langle G, \emptyset, true, \emptyset \rangle_1
$.

The rules of a program $\Prog$ are applied to exhaustion on this initial program state. 
A rule is {\em applicable}, if 
its head constraints are matched by constraints in the current CHR store
one-by-one 
and if, under this
matching, the guard of the rule is implied by the built-in constraints in the
goal.
Any of the applicable rules can
be applied, and the application cannot be undone, it is committed-choice (in
contrast to Prolog).
When a simplification rule is applied, the matched constraints in the current
CHR store are replaced by the body of the rule; when a propagation rule is applied,
the body of the rule is added to the goal without removing any constraints.

\begin{figure}
\begin{center}
\fbox{
\begin{minipage}{0.9\textwidth}
\noindent \textbf{1. Solve} 
$
        \langle \{c\} \uplus G, S, B, T \rangle_n \rightarrowtail_{solve}
        \langle G, S, c \wedge B, T \rangle_n
$
where $c$ is a built-in constraint. \\
\rule{\textwidth}{0.5pt}
\noindent \textbf{2. Introduce} 
$
        \langle \{c\} \uplus G, S, B, T \rangle_n \rightarrowtail_{introduce}
        \langle G, \{\num{c}{n}\} \uplus S, B, T \rangle_{(n+1)}
$
where $c$ is a CHR constraint. \\
\rule{\textwidth}{0.5pt}
\noindent \textbf{3. Apply} 
$
         \langle G, H_1 \uplus H_2 \uplus S, B, T \rangle_n \rightarrowtail_{apply}
         \langle C \uplus G, H_1 \uplus S, \theta \wedge B, T' \rangle_n
$
where there exists a (renamed apart) rule in $\Prog$ of the form 
$$
        r\ @\ H'_1\ \backslash\ H'_2 \iff g\ |\ C
$$
and a matching substitution $\theta$ such that
$chr(H_1) = \theta(H'_1)$, $chr(H_2) = \theta(H'_2)$
and $\BT \models B \rightarrow \bar{\exists}_B (\theta \wedge g)$.
In the result $T' = T \cup
\{id(H_1)\plusplus id(H_2)\plusplus [r]\}$.
It should hold that $T' \neq T$.
\end{minipage}
}
\end{center}
\caption{The transition rules of the operational semantics of CHR}\label{fig:chr:semantics}
\end{figure}

\subsection{Implementation of CHR}\label{sec:chr:implementation}

This high-level description of the operational semantics of CHR leaves two main
sources of non-determinism: the order in which constraints of a query are
processed and the order in which rules are applied.\footnote{The nondeterminism due to 
the wake-up order of delayed constraints and multiple matches for the same rule
are of no relevance for the programs discussed here.} As in Prolog, almost all CHR
implementations execute queries from left to right and apply rules top-down in
the textual order of the program. This behavior has
been formalized in the so-called refined semantics that was also proven to be a
concretization of the standard operational semantics \cite{duck:refined}.

In this refined semantics of actual implementations, a CHR constraint in a query
can be understood as a procedure that goes efficiently through the rules of the program
in the order they are written,
and when it matches a head constraint of a rule, it will look for the other,
{\em partner
constraints} of the head in the {\em constraint store} and check the guard until an
applicable rule is found. We consider such a constraint to be {\em active}. If the
active constraint has not been removed after trying all rules, it will be put
into the constraint store. Constraints from the store will be reconsidered
(woken) if newly added built-in constraints constrain variables of the
constraint, because then rules may become applicable if their guards
are now implied.

The refined operational semantics is implemented by all major CHR systems,
among which the K.U.Leuven CHR system \cite{tom:kulchr}. This system is
currently available in three different Prolog systems (hProlog \cite{hprolog},
SWI-Prolog \cite{swiprolog} and XSB) and it serves as the basis of our
integration with tabled execution in this paper.

The K.U.Leuven CHR system 
\cite{tom:kulchr} is based on the general compilation schema of CHR by
Holzbaur \cite{christian:system}. For this paper (Section
\ref{sec:encoding}) it is relevant to know that the CHR
constraint store is implemented as a global updatable term, containing
identified constraints, in this context also called {\em suspended constraints}, grouped by
their functor. Each suspended constraint $\num{c}{i}$ is represented as a suspension term,
including the following information:
\begin{itemize}
\item The constraint $c$ itself.
\item The constraint identifier $i$.
\item The continuation goal, executed on reactivation. This
      goal contains the suspension itself as an argument and it is in fact
      a cyclic term.
\item The part of the propagation history $T$ containing for each propagation rule the tuple
      of identifiers of other constraints that this constraint has interacted
      with.
\end{itemize}
Variables involved in the suspended constraints behave as indexes
into the global store: they have the suspensions attached to them as attributes.
Because we aim towards a light-weight integration of CHR with tabled Logic
Programming, we do not question these established representation properties,
but consider them as something to cope with.

We refer the interest reader to \cite{tom:phd} for more details on CHR
implementation.

\subsection{CHR for Constraint Solving}\label{sec:chr:cp}

The CHR language is intended as a language for implementing {\em constraint
solvers}. 

A CHR program $\Prog$ is a constraint solver for the constraint domain
$\CT_\Prog$ whose constraint symbols $\Pi_\Prog$ are the CHR and built-in constraint
symbols. The constraint theory $\mathcal{T}_\Prog$ of the program consists
of the built-in constraint theory together with the declarative meaning
of the CHR rules. The declarative meaning of a simplification rule of the
form $H_r~\texttt{<=>}~G~\texttt{|}~B$ is:
\[\forall \bar{x}. \exists \bar{y} . G \rightarrow (H \leftrightarrow \exists \bar{z} . B)\]
where $\bar{x} = \vars{H} \cup$, $\bar{y} = \vars{G} \setminus \vars{H}$
and $\bar{z} = \vars{B} \setminus (\vars{H} \cup \vars{G})$.  Similarly,
the declarative meaning of a propagation rule of the form
$H~\texttt{==>}~G~\texttt{|}~B$ is:
\[\forall \bar{x} . \exists y . G \rightarrow (H \rightarrow \exists \bar{z} B)).\]
Value sets are not explicitly defined by the CHR program, but they exist
implicitly in the intention of the programmer.

See \cite{thom:book} for an extensive treatment of
CHR for writing constraint solvers.

\section{The TCHR Framework}\label{sec:tabling}

The main challenge of introducing CHR in XSB is the integration of CHR
constraint solvers with the backward chaining fixpoint computation of SLG
resolution according to the SLG$^\CT$ semantics of the previous section.

A similar integration problem has been solved in \cite{bao:tclp}, which
describes a framework for constraint solvers written with attributed variables
for XSB. The name Tabled Constraint Logic Programming (TCLP) is coined in
that publication, though it is not formulated in terms of SLG$^\CT$ resolution.
Porting CHR to XSB was already there recognized as important future work. 

CHR is much more convenient for developing constraint solvers than attributed
variables, because of its high-level nature. This advantage should be carried
over to the tabled context, making tabled CHR a more convenient paradigm
than TCLP with attributed variables.
Indeed, we will show how the internal details
presented in the current section can be hidden from the user.


In \cite{bao:tclp} the general TCLP framework specifies three operations to
control the tabling of constraints: call abstraction, entailment checking
of answers and answer projection. These operations correspond with the
optimization to {\em Query Projection}, the projection in {\em Answer
Projection} and the compaction of the $\ans(G;C)$ set. It is left
to the constraint solver programmer to implement these operations for his particular solver.

In the following we formulate these operations in terms of CHR. The operations
are covered in significant detail as the actual CHR implementation and the
encoding of the global CHR constraint store are taken into account.


\subsection{General Scheme of the TCHR Implementation}\label{sec:tabling:schema}

The objective of TCHR is to implement \slgc\ semantics for an arbitrary
constraint domain $\CT$ which is implemented as a CHR constraint solver. For
this purpose we have both an \slg{\Herbrand} implementation\footnote{SLG is a special case of \slgc\ where $\CT$ is the Herbrand constraint domain $\Herbrand$.}, i.e. the
SLG implementation of XSB, and an SLD$^\CT$ implementation, i.e. the CHR
implementation of XSB, at our disposal.

Hence, we aim for the simplest and least intrusive solution. That is:
\begin{enumerate}
\item We use the {\em unmodified} CHR implementation for constraint solving.
\item We use the {\em unmodified} SLG implementation for tabled execution.
\item At the intersection of point 1 and point 2 we transform back and forth between the CHR $\CT$ constraints and
      an encoding of these as $\Herbrand$ constraints.
\end{enumerate}

The advantages to this lightweight approach are twofold. Firstly, it
is straightforward to realize the full expressivity of \slgc\ within an existing
system. Secondly, it does not affect existing programs or their performance.
On the downside we note that TCHR performance and, in particular, constant
factors involved are not optimal. However, CHR on its own does not aim towards
performance in the first place, but rather towards being a highly expressive
formalism for experimenting with new constraint solvers. Similarly, we see
the TCHR framework as a highly expressive prototyping system for exploring
new applications.  It does offer some high-level means to affect performance,
and when the resulting performance is simply not good enough, one may decide
to reimplement the established high-level approach in a lower-level language.

Now we look at our solution in more detail. As points 1 and 2 leave the
system untouched, we only have to consider implementing point 3, translating
between constraint encodings. 

First let us consider the different kinds of nodes used in \slgc. Of the
tree nodes, only the root and answer nodes are manifestly represented
by \slg{\Herbrand} implementations like XSB, in respectively call and
answer tables.  Hence these two nodes require the constraint store to be in
$\Herbrand$ encoding form. The other two nodes, the goal and body nodes,
are implicit in the execution mechanism. So here we are free to use the form
that suits us best.

With these formats for the nodes in mind, we consider one by
one the different resolution rules:
\begin{description}
\item[Clause Resolution] The rule is depicted again below with each constraint annotated with its  
type of encoding: $\Herbrand$ for Herbrand encoding and $\chr$ for natural CHR encoding. 
The constraint store $C$ is initially Herbrand encoded in the root node and has to be decoded
into its natural CHR form for solving $C \wedge \theta \wedge D^i$ with the CHR solver. The CHR
solver either fails, if the conjunction is not satisfiable, or returns a simplified form of the conjunction.

{\small
\renewcommand{\arraystretch}{1.3}
\begin{tabular}{@{\hspace{1mm}}r@{}l@{\hspace{-5pt}}l@{}}
$\texttt{root}(G;C_\Herbrand)$ &

$
\left\{\begin{array}{c}
\texttt{body}(G;B_1^1,\ldots,B_{k_1}^1;C_\chr \wedge \theta \wedge D_\chr^1) \\
\vdots \\
\texttt{body}(G;B_1^l,\ldots,B_{k_l}^l;C_\chr \wedge \theta \wedge D_\chr^l) \\
\end{array}\right.
$ 

& 
 
\begin{minipage}{0.3\textwidth}
$\forall i. 0 < i \leq l$ such that \\
$G' \rightarrow D_\chr^i, B_1^i,\ldots,B_{k_i}^i$ \\
and $\theta \equiv (G = G')$ \\
and $C_\chr \wedge \theta \wedge D_\chr^i$ is satisfiable \\
\end{minipage}
\\
\end{tabular}
}

\item[Optimized Query Projection] 
        This rule directly starts with a constraint store in the natural
        CHR constraint form, and projects it onto the first literal and
        subsequently generalizes it. A CHR program does not normally come
        with such a combined projection \& generalization operation, so one
        will have to be supplied by the TCHR framework: the {\em call
        abstraction}. Section \ref{sec:abstraction} discusses what kind of
        generic projection operation the TCHR framework implements.

{\small
\renewcommand{\arraystretch}{1.3}
\begin{tabular}{@{\hspace{1mm}}r@{}l@{\hspace{-5pt}}l@{}}
$\texttt{body}(G;B_1,\ldots,B_k;C_\chr)$ &

$
\left\{\begin{array}{c}
\texttt{goal}(G;B_1,C_\chr^1;B_2\ldots,B_k;C_\chr) \\
\vdots \\
\texttt{goal}(G;B_1,C_\chr^l;B_2\ldots,B^k;C_\chr) \\
\end{array}\right.
$ 

& 
 
\begin{minipage}{0.3\textwidth}
$\CT \models \project{B_1}C \rightarrow C_1 \vee \ldots \vee C_l$ \\
for some $C_1,\ldots,C_l$
\end{minipage}
\end{tabular}
}

\item[Answer Propagation] 
        The answers consumed by this rule have to be decoded from Herbrand form
        for the implication check and the satisfiability check. 
        A CHR program does not normally come with an implication check, so
        one will have to be supplied here by the TCHR framework. This is covered together with
        call abstraction in Section \ref{sec:abstraction}.

{\small
\renewcommand{\arraystretch}{1.3}
\begin{tabular}{@{\hspace{1mm}}r@{}l@{\hspace{-5pt}}l@{}}
$\texttt{goal}(G;B_1,C_\chr^1;B_2\ldots,B_k;C_\chr)$ &

$
\left\{\begin{array}{c}
\texttt{body}(G;B_2,\ldots,B_k;C_\chr \wedge \theta \wedge A_\chr^1) \\
\vdots \\
\texttt{body}(G;B_2,\ldots,B_k;C_\chr \wedge \theta \wedge A_\chr^l) \\
\end{array}\right.
$ 

& 
 
\begin{minipage}{0.3\textwidth}
$\forall A_\Herbrand^i \in \ans(B',C_\Herbrand')$ \\
where $\theta \equiv (B' = B_1)$ \\
and $C_\chr^1 \wedge \theta \leq_{\CT} C'_\chr$ \\
and $C_\chr \wedge \theta \wedge A_\Herbrand^i$ is satisfiable
\end{minipage}
\\
\end{tabular}
}
\item[Answer Projection]

Again a projection is performed on the CHR constraint representation. This
instance of of projection we call {\em answer projection}. Like
answer projection, it is to be supplied by the framework. In Section
\ref{sec:projection} the details of this operation within the framework are 
elaborated.

{\small
\renewcommand{\arraystretch}{1.3}
\begin{tabular}{@{\hspace{1mm}}r@{}l@{\hspace{-5pt}}l@{}}
$\texttt{body}(G;\Box;C_\chr)$ &

$
\left\{\begin{array}{c}
\texttt{ans}(G;A_\Herbrand^1) \\
\vdots \\
\texttt{ans}(G;A_\Herbrand^l) \\
\end{array}\right.
$ 

& 
 
\begin{minipage}{0.3\textwidth}
for all $A_\chr^i \in \project{G}C$ 
\end{minipage}
\\
\end{tabular}
}
\end{description}

Having established what new operations and mappings to include in the
framework, we should consider how these are to be incorporated into the
existing \slg{\Herbrand}\ system XSB.
Recall that we did intend not to modify the system to incorporate
our encoding/decoding and projection operations in order to keep the integration
light-weight. Neither do we want to encumber the programmer with this tedious
and rather low-level task. Instead we propose an automatic source-to-source
transformation based on a simple declaration to introduce these
operations.

The source-to-source transformation maps the SLG$^\CT$ program $P$ onto the
\slg{\Herbrand}\ program $P'$. 
In the mapping every predicate
$p/n \in P$ is considered independently, and mapped onto three predicates
$p/n, \mathit{tabled\_p}/(n+2), \mathit{original\_p}/n \in P'$.
$\transfo$ maps the SLG$^\CT$ program
$P$ onto the \slg{\Herbrand}\ program $P' = \transfo(P)$. 

We outline the high-level transformation for a single predicate \texttt{p/2}:
\begin{Verbatim}[commandchars=\\\{\}]
        :- table p/2.

        p(X,Y) :- \emph{Body}. 
\end{Verbatim}

The three resulting predicates are:
\begin{Verbatim}[commandchars=\\\{\}]
        p(X,Y) :- 
          encode_store(CurrentStoreEncoding),
          call_abstraction([X,Y],CurrentStoreEncoding,AbstractStoreEncoding),
          empty_store,
          tabled_p(X,Y,AbstractStoreEncoding,AnswerStoreEncoding),
          decode_store(CurrentStoreEncoding),
          decode_store(AnswerStoreEncoding). 

        :- table tabled_p/4.

        tabled_p(X,Y,StoreEncoding,NStoreEncoding) :-
          decode_store(StoreEncoding),
          original_p(X,Y),
          encode_store(StoreEncoding1),
          empty_store,
          project([X,Y],StoreEncoding1,NStoreEncoding).

        original_p(X,Y) :- \emph{Body}.
\end{Verbatim}

The new predicate \texttt{p/2} is a front to the actual tabled predicate
\texttt{tabled\_p/4}. This front allows the predicate to be called with the
old calling convention where the constraint store is implicit, i.e. in the
natural CHR form. Thanks to this front the transformation is modular: we
do not have to modify any existing calls to the predicate, either in other
predicates' bodies, its own body {\em Body} or in queries.  The auxiliary
predicate \texttt{encode\_store/1} returns a Herbrand encoding of the
current (implicit) constraint store and the \texttt{call\_abstraction/3} predicate
projects the Herbrand encoded store onto the call arguments. Then the
implicit constraint store is emptied with \texttt{empty\_store/0} so as not
to interfere with the tabled call, which has the Herbrand encoded stores as
manifest answers. Finally, the predicate \texttt{decode\_store/1} decodes
the Herbrand encoding and adds the resulting CHR constraint store to the implicit store.
In \texttt{p/2} this predicate is called twice: first to restore the current constraint
store and then to add to it the answer constraint store of the tabled call.

The \texttt{tabled\_p/4} predicate is the tabled predicate. In its body the
encoded input store is decoded again, then the original predicate code
\texttt{original\_p/2} is run, the resulting store is encoded again and
projected onto the call arguments. Again the implicit store is emptied so
as not to interfere with the caller.

Note that this is only a high-level outline of the mapping. In practice the
scheme is specialized for the concrete operations. This is discusses later
as we discuss each of the operations in detail.



The above transformation of both predicates and queries can be fully transparent. All
the user has to do is to indicate what predicates have to be tabled,
i.e. add a declaration of the form
\begin{Verbatim}[commandchars=\\\{\}]
                :- table_chr p(_,chr) with \emph{Options}.

                p(X,Y) :- ...
\end{Verbatim}
meaning that the predicate \texttt{p/2} should be tabled, its first
argument is an ordinary Prolog term and its second argument is a CHR
constraint variable. An (optional) list of additional options \texttt{{\em Options}} 
may be provided to control the transformation:
\begin{description}
\item[\texttt{encoding({\em EncodingType})}] \ \\
        Section \ref{sec:encoding} studies two alternative
        encodings of the Herbrand constraint store. This option allows the user to choose between
        them.
\item[\texttt{projection(\emph{PredName})}] \ \\
        The projection applied in the {\em Answer Projection} rule is addressed
        in Section \ref{sec:projection}. This projection
        is realized as a call to a projection predicate that reduces the constraint
        store to its projected form.
\item[\texttt{canonical\_form(\emph{PredName})}] 
\item[\texttt{answer\_combination(\emph{PredName})}] \ \\

        These two options relate to optimizations of the answer
        set, based on Definition \ref{def:tabling:slgc:entailment}
        and a novel generalization of this principle. It is discussed
        in Section \ref{sec:asopt}.
\end{description}

\begin{figure}[h]
\begin{center}
    \includegraphics[width=\textwidth]{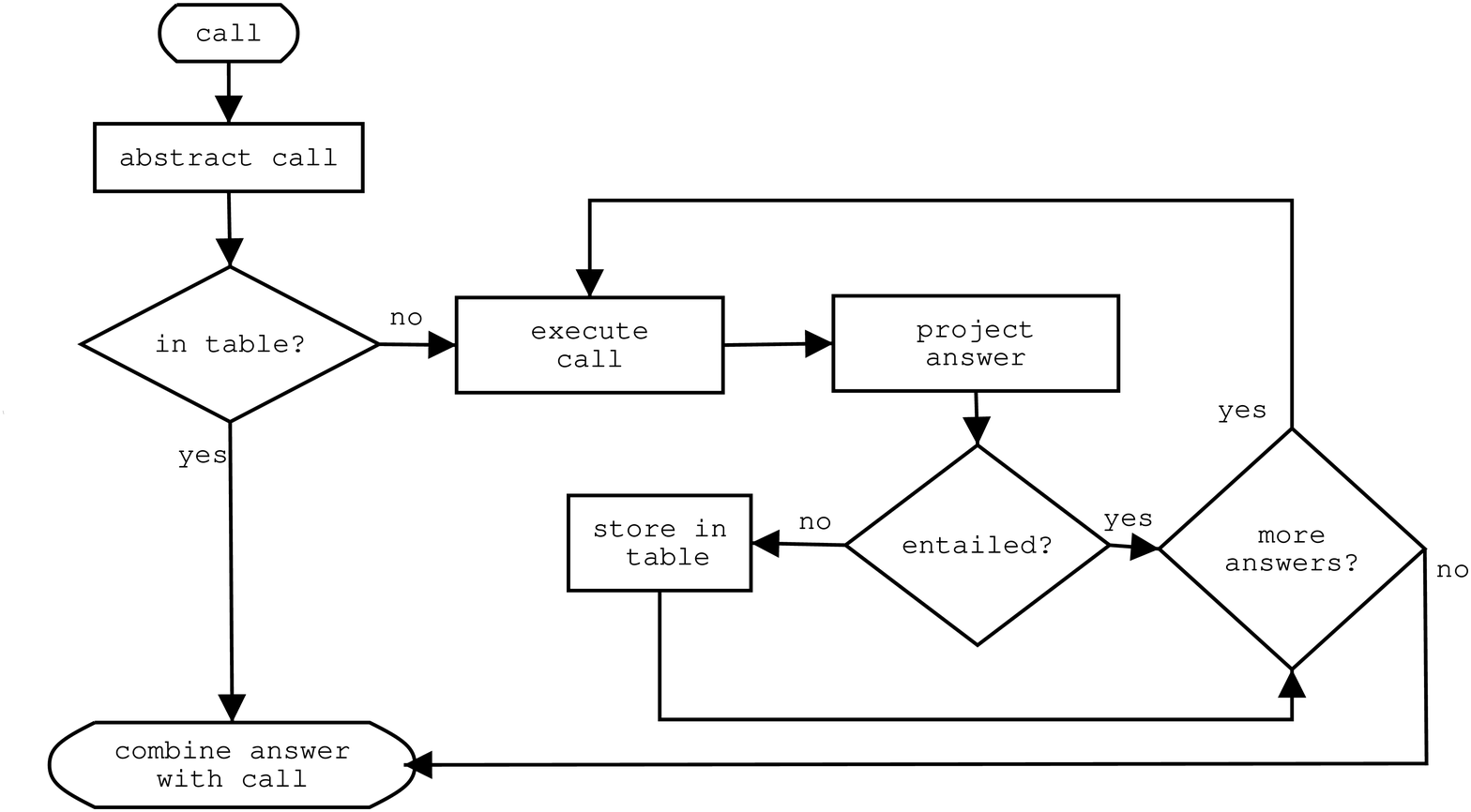}
\end{center}
    \caption{Tabled call flowchart\label{fig:flowchart}}
\end{figure}

Figure \ref{fig:flowchart} summarizes the different steps in handling a call to a tabled predicate. 

\section{Herbrand Constraint Store Encodings}\label{sec:encoding}

In this section we present two alternative Herbrand constraint store encodings. 
An encoding must have the following properties:
\begin{itemize}
\item 
The encoding has to be suitable for passing it as an argument in a predicate
and for storing it in an answer table. 
\item 
It should be possible to convert from the natural CHR constraint form
(see Section \ref{sec:chr:implementation}) and back, for insertion into the call table and retrieval from
the answer table.
\end{itemize}

%

The essential aspects of the ordinary CHR constraint store implementation have
been covered in Section \ref{sec:chr:implementation}.  Two different Herbrand constraint 
store encodings that are based on this ordinary form have been
explored: the {\em suspension} encoding and the {\em goal}
encoding. The former is based on state copying and the latter on recomputation.
A discussion of their respective merits and weaknesses as well
as an evaluation follow in Sections \ref{sec:encoding:suspension} and \ref{sec:encoding:goal} respectively.

One implicit aspect of CHR execution under the refined operational semantics
is the order in which constraints are processed. Ordering information
is not maintained explicitly. Without any additional support, it is not
straightforward to maintain this ordering information for tabled constraints.
However, in the spirit of tabling, the declarative meaning of a program rather
than its operational behavior is of importance. For that reason we shall
not attempt to realize the ordering of the refined operational semantics.
From the user's point of view, the CHR constraints behave according
to the theoretical operational semantics and no assumptions should be made
about ordering.


\subsection{Suspension Encoding}\label{sec:encoding:suspension}

This encoding aims at keeping the tabled encoding as close as possible to the
ordinary form. The essential issue is to retain the propagation history
of the constraints. In that way no unnecessary re-firing of propagation
rules occurs after the constraints have been retrieved from the table.

However, it is not possible to just store the ordinary constraint suspensions
in the table as they are. Fortunately, attributed variables themselves can
be stored in tables (see \cite{bao:attvars}), but two other aspects have to
be taken into account.  Firstly, these suspensions are cyclic terms that
the tables cannot handle.  This can be dealt with by breaking the cycles
upon encoding and resetting them during decoding. Secondly, the constraint
identifiers have to be replaced by fresh ones during decoding, as multiple
calls would otherwise create multiple copies of the same constraints all
with identical identifiers. Finally, after decoding, the constraints have
to be activated again in order to solve them together with the already
present constraints. This is done by simply calling their continuation goals.

\begin{example}
Let us consider the following program:
\begin{Verbatim}
        :- constraint a/0, b/0.
        r1 @ a ==> b.

        :- chr_table p.
        p :- a.
\end{Verbatim}
and the query \texttt{?- p.} After having fired rule \texttt{r1}, the suspension of an \texttt{a} constraint 
looks like:
\[ Sa = \texttt{suspension(42,reactivate\_a($Sa$),[r1-[42]])}\]
where \texttt{42} is the identifier, \texttt{reactivate\_a($Sa$)} is the continuation
goal and \texttt{[r1-[42]]} is the propagation history, which has recorded that rule \texttt{r1}
has fired with only the constraint itself. The other suspension in the store would be for a \texttt{b}
constraint:
\[ Sb = \texttt{suspension(43,reactivate\_b($Sb$),[])}\]

The suspension encoding for this store of two constraints would look like:
\begin{Verbatim}
        [ Sa / ID1 / suspension(ID1,reactivate_a(Sa),[[r1-[ID1]]])
        , Sb / ID2 / suspension(ID2,reactivate_b(Sb),[])]
\end{Verbatim}

Upon decoding we simply unify \texttt{ID1} and \texttt{ID2} with fresh
identifiers, and \texttt{S1} and \texttt{S2} with their corresponding
suspension terms. The resulting well-formed suspension terms are placed in
the implicit CHR constraint store and finally the continuation goals of both
suspensions are called.
\end{example}

\subsection{Goal Encoding}\label{sec:encoding:goal}

The goal encoding aims at keeping the information in the table in as
simple a form as possible: for each suspended constraint only the goal to
impose this constraint is retained in the table. It is easy to create this goal
from a suspension and easy to merge this goal back into another constraint
store: it needs only to be called.

Whenever it is necessary the goal creates a suspension with a fresh
unique identifier and inserts it into the constraint store. 

The only information that is lost in this encoding is the propagation
history. This may lead to multiple propagations for the same combination
of head constraints. For this to be sound, a further restriction on the CHR
rules is required: they should behave according to set semantics, i.e. the
presence of multiple identical constraints should not lead to different
answers modulo identical constraints.

\begin{example}
The goal encoding of the above example is:
\begin{Verbatim}
        [a, b]
\end{Verbatim}
and the decoding procedure simply calls \texttt{a} and \texttt{b}.
\end{example}

\subsection{Evaluation}\label{sec:encoding:evaluation} 

To measure the relative performance of the two presented encodings, consider
the following two programs:
\begin{center}
\begin{tabular}{lll}
\begin{minipage}{.48\textwidth}
\begin{Verbatim}[frame=single,framesep=1mm,label=\textbf{prop}]
:- constraints a/1.                
  a(0) <=> true.                     
  a(N) ==> N > 0 
       | M is N - 1, a(M).                             

  p(N) :- a(N).                      
\end{Verbatim}
\end{minipage} & \hspace{-6mm} &
\begin{minipage}{.48\textwidth}
\begin{Verbatim}[frame=single,framesep=1mm,label=\textbf{simp}]
:- constraints a/1, b/1.
  b(0) <=> true.
  b(N) <=> N > 0 
       | a(N), M is N - 1, b(M).

  p(N) :- b(N).
\end{Verbatim}
\end{minipage} \\
\end{tabular}
\end{center}
For both programs the predicate \texttt{p(N)} puts the constraints
\texttt{a(1)}...\texttt{a(N)} in the constraint store. The \texttt{prop}
program uses a propagation rule to achieve this while the \texttt{simp}
program uses an auxiliary constraint \texttt{b/1}. The non-tabled version
of the query \texttt{p(N)} has time complexity $\mathcal{O}(N)$
for both the \texttt{simp} and the \texttt{prop} program.

The two possible encodings for the answer constraint store can be
specified in the tabling declaration as follows:
\begin{Verbatim}
        :- table_chr p(_) with [encoding(suspension)].
\end{Verbatim}
and 
\begin{Verbatim}
        :- table_chr p(_) with [encoding(goal)].
\end{Verbatim}

\begin{table}
\caption{\label{tab:encoding}Evaluation of the two tabled store encodings.}
\begin{center}
\begin{tabular}{|l||r|r|r|r|r|r|}
\cline{1-7}
        &  \multicolumn{2}{c|}{no tabling} & \multicolumn{2}{c|}{\texttt{encoding(suspension)}} & \multicolumn{2}{c|}{\texttt{encoding(goal)}} \\
program & runtime  & space     &     runtime  & space         & runtime  &       space \\
\cline{1-7}
prop  & 10 & 0 & 150   & 2,153,100     & 1,739 &       270,700 \\
\cline{1-7}
simp  & 10 & 0 & 109   & 1,829,100     &    89 &       270,700 \\
\cline{1-7}
\end{tabular}
\end{center}
\end{table}

Table \ref{tab:encoding} gives the results for the query \texttt{p(400)}, both
untabled and tabled using the two encodings: runtime in milliseconds and space
usage of the tables in bytes. For both programs the answer table contains the
constraint store with the 400 \texttt{a/1} constraints. 

Most of the space overhead is due to the difference in encoding:
a suspension contains more information than a simple call. However, the
difference is only a constant factor. The only part of a suspension in
general that can have a size greater than $\mathcal{O}(1)$ is the propagation
history. In the \texttt{prop} program every \texttt{a/1} constraint's history
is limited to remembering that the propagation rule has been used once. For
the \texttt{simp} program the propagation history is always empty.

The runtime of the \texttt{prop} version with the suspension encoding is
considerably better than that of the version with the goal encoding. In
fact, there is a complexity difference. When the answer is retrieved from the
table for the suspension encoding, the propagation history prevents
re-propagation. Hence, answer retrieval takes $\mathcal{O}(N)$ time. However, for the goal 
encoding every constraint \texttt{a(I)} from the
answer will start propagating and the complexity of answer retrieval becomes
$\mathcal{O}(N^2)$.

On the other hand, for \texttt{simp} the propagation history plays no role. The
runtime overhead is mostly due to the additional overhead of the pre- and
post-processing of the suspension encoding as opposed to the simpler
form of the goal encoding.
In comparison, without tabling the query takes only 10 milliseconds for both
programs.
\section{Call Abstraction}\label{sec:abstraction}

In the call abstraction operation we combine the projection and generalization
operations of the {\em Optimized Query Projection} rule in the \slgc\
semantics. 

The idea of both steps is to reduce the number of distinct \slgc\ trees, and
hence the number of tables. When a predicate is called with many different
call patterns, a table is generated for each such call pattern. Thus it is
possible that the information for one strongly constrained call is present
many times in tables for different less constrained call patterns. This
duplication in the tables can be avoided by using call abstraction to obtain
a smaller set of call patterns.
 
The projection reduces the context of the predicate call, i.e. the constraint
store,
to the constraints relevant for the call. In this way, two calls to
\texttt{p(X)}, respectively with constraint stores
\texttt{\{X > 5, Y > 7\}} and \texttt{\{X > 5, Z < 3\}} both yield the same
projected call store \texttt{\{ X > 5 \}}. The subsequent generalization step
goes even further, e.g. by relaxing bounds to the reference value 0 
both constraint stores \texttt{\{X > 5\}} and \texttt{\{X > 10\}} become
\texttt{\{X > 0\}}.
Hence, call abstraction effectively is a means to control the number of tables. 
At the level of \slg{\Herbrand}, call abstraction means not passing certain
bindings to the call.  For example, \texttt{p(q,A)} can be abstracted to
\texttt{p(Q,A)}. This goal has then to be followed by \texttt{Q = q} to
ensure that only the appropriate bindings for \texttt{A} are retained.

For \slgc, call abstraction can be generalized from
bindings to constraints: abstraction means removing some of the constraints
on the arguments. Consider for example the call \texttt{p(Q,A)} with
constraint \texttt{Q =< N} on \texttt{Q}. This call can be abstracted to
\texttt{p(Q',A)}, followed by \texttt{Q'=Q} to reintroduce the constraint.

Abstraction is particularly useful for those constraint solvers for which the
number of constraints on a variable can be much larger than the number of
different bindings for that variable. Consider for example a finite domain
constraint solver with constraint \texttt{domain/2}, where the first argument
is a variable and the second argument the set of its possible values. If the
variable has a domain of size $n$ (i.e. it contains $n$ different values), the
variable can take as many as $2^n$ different \texttt{domain/2} constraints,
one for each subset of values. Thus many different tables would be needed
to cover every possible call pattern.

Varying degrees of abstraction are possible, depending on the particular
constraint system or application. Full constraint abstraction, i.e. the
removal of all constraints from the call, is generally the only option for CHR,
for the following reasons:
\begin{itemize}
\item CHR rules do not require constraints to be on variables.
      They can be exclusively on ground terms or atoms as well. This is useful
      for various reasons. By encoding constraint variables as ground terms,
      particular solving algorithms can be used more conveniently or efficiently, e.g. 
      the equation solving algorithm union-find has optimal time-complexity when
      using ground elements \cite{tom:unionfind:pearl}.  

      It is not straightforward to automatically define abstraction for
      ground terms as these are not necessarily passed in as arguments but
      can just as well be created inside the call. Hence there is no explicit
      link with the call environment, while such a link is needed for call
      abstraction. As such, only ``no abstraction'' or full constraint
      abstraction seem suitable for CHR.
\item Full constraint abstraction is preferable when the previously mentioned 
      table blow-up is likely. 
\item In order to reuse existing answers, existing calls are considered in the
      {\em Answer Propagation} rule. These previous calls are compared to the
      new call using the implication check $\leq_\CT$. Unfortunately, such an implication
      check does not come with the CHR solver. A special case of this subsumption-based
      tabling is where $\leq_\CT$ is taken to be $\leftrightarrow$, i.e. equivalence-based,
      or variant-based in \slg{\Herbrand} terminology, tabling. Unfortunately, even
      establishing the equivalence of constraint stores is not directly supported by CHR
      solvers.
      
      However, if the call constraint store is empty, i.e. $\true$, this
      problem disappears: $\true$ implies $\true$ independent of the constraint
      domain. 
\end{itemize}

Moreover, it may be costly to sort out what constraints should be passed
in to the call or abstracted away. Hence often full abstraction is cheaper
than partial abstraction. For instance, consider a typical propagation-based
finite domain constraint solver with binary constraints only. The constraint
graph for a number of such finite domain constraints has a node for every
variable involved in a constraint and an edge between variables involved
in the same constraint. Any additional constraint imposed on a variable in
a component of the graph may affect the domain of all other variables in
the same component. Hence, call abstraction on a subset of the variables
involves a costly transitive closure of reachability in the constraint graph.

Let us now revisit the transformation scheme of Section
\ref{sec:tabling:schema} for a predicate \texttt{p/2}, and specialize it for full
call abstraction:
\begin{Verbatim}[commandchars=\\\{\}]
        p(X,Y) :- 
          encode_store(StoreEncoding),
          empty_store,
          tabled_p(X,Y1,NStoreEncoding),
          decode_store(StoreEncoding),
          decode_store(NStoreEncoding),
          Y1 = Y. 

        :- table tabled_p/3.

        tabled_p(X,Y,NStoreEncoding) :-
          original_p(X,Y),
          encode_store(StoreEncoding1),
          empty_store,
          project([X,Y],StoreEncoding1,NStoreEncoding).

        original_p(X,Y) :- \emph{Body}.
\end{Verbatim}

As we know that the call-abstracted constraint store is empty, we no
longer need to pass it as an argument to the \texttt{tabled\_p} predicate
and decode it there. The only effect of the call abstraction is then to
replace the constraint variable Y with a fresh variable \texttt{Y1}. This
is necessary to prevent any constraints on \texttt{Y} from being reachable
through attributes on \texttt{Y}. The unification \texttt{Y=Y1} at the end of \texttt{p/2}
is then a specialization of the substitution $\theta \equiv (p(X,Y1) = p(X,Y))$ that
appears in the {\em Answer Propagation} rule.


%

\section{Answer Projection}\label{sec:projection}

In most constraint domains, the same logical answer can be represented in many
different ways. For example, consider the predicate \texttt{p/1}.
\begin{Verbatim}[commandchars=\\\{\}]
	p(X) :- X > 5.
	p(X) :- X > 5, Y > 0.
\end{Verbatim}

Both $X > 5$ and $X > 5 /\ Y > 0$ represent the same answer to the call
\texttt{p($X$)} concerning $X$.  Constraints that do not relate to the call
arguments, like $Y > 0$, are meaningless outside of the call. The local
variable $Y$ is existentially quantified, and cannot be further constrained to
introduce unsatisfiability at a later stage.

It is the purpose of projection to restrict constraints to a set of variables
of interest, and to eliminate other variables as much as possible. In our
setting, the variables of interest are the call arguments.  For projection to
be sound, already present but not yet detected unsatisfiability should not be
removed. A sufficient, but not necessary condition is for the constraint system
to be complete, i.e. unsatisfiability is detected immediately.

Projection is important in the context of tabling, because it may give
logically equivalent answers the same syntactical form. When two answers have
the same syntactical form, the are recognized as duplicates and only one is
retained in the table. A vital application of projection is when a predicate
with an infinite number of different answers may be turned into one with just a
finite number of answers by discarding the constraints on local variables.

\begin{example}\label{ex:leq}
Consider this program:
\begin{Verbatim}
	path(From,To,X) :- 
	        edge(From,To,X).
	path(From,To,X) :- 
	        path(From,Between,X), path(Between,To,X). 
	
	edge(a,a,X) :-
	        leq(X,Y),
	        leq(Y,1).
	
	leq(X,X)             <=> true.
	leq(X,Y) \ leq(Y,X)  <=> X = Y.
	leq(X,Y) \ leq(X,Y)  <=> true.
	leq(X,Y) , leq(Y,Z)  ==> leq(X,Z).
\end{Verbatim}
It defines a \texttt{path/3} predicate that expresses reachability in a graph
represented by \texttt{edge/3} predicates. The first two arguments of both
predicates are edges (origin and destination) and the third is a constraint
variable. Along every edge in the graph some additional constraints may be
imposed on this variable. In our example, the graph consists of a single loop
from edge \texttt{a} to itself. This loop imposes two less-than-or-equal-to
constraints: \texttt{leq(X,Y), leq(Y,1)}. The variable \texttt{Y} is a local
variable and the fourth rule for \texttt{leq/2} derives that \texttt{leq(X,1)}
also holds.

The query \texttt{?- path(A,B,X)} determines the different paths.
There are an infinite number of paths in our simple graph, one for each
non-zero integer $n$. A path for $n$ takes the loop $n$ times. For every
time the loop is taken a new variable \texttt{Y}$_i$ is created and two more
constraints \texttt{leq(X,Y$_i$)} and \texttt{leq(Y$_i$,1)} are added. Through
the propagation rule also an \texttt{leq(X,1)} is added for each time the
loop is taken. The second simpagation rule however removes all but one copy
of this last constraint.

Even though there are an infinite number of answers, the constraints involving
the local variables \texttt{Y$_i$} are of no interest and only the single
\texttt{leq(X,1)} is relevant.
\end{example}

In general constraint projection onto a set of variables transforms a
constraint store into another constraint store in which only variables of the
given set are involved. The form of the resulting constraint store strongly
depends on the particular constraint solver and its computation may involve
arbitrary analysis of the original constraint store.

We propose what we believe is an elegant CHR-based approach to projection. It
consists of a compact and high level notation. 

The user declares the use of the CHR-based approach to projection as follows:
\begin{Verbatim}[commandchars=\\\{\},codes={\catcode`$=3}]
	        :- table_chr p(_,chr) with [projection($PredName$)].
\end{Verbatim}
and implements the projection as a number of CHR rules that involve the
special \texttt{$PredName$/1} constraint. This constraint has as its argument
the set of variables to project on.

The source-to-source transformation generates the predicate \texttt{tabled\_p}
based on this declaration:
\begin{Verbatim}[codes={\catcode`$=3}]
        :- table tabled_p/3.

        tabled_p(X,Y,NStoreEncoding) :-
          original_p(X,Y),
          $PredName$([X,Y]),
          encode_store(NStoreEncoding),
          empty_store.
\end{Verbatim}
When no projection operation is supplied, the default action is
to return the constraint store unmodified.


To implement the projection simpagation rules can be used to decide
what constraints to remove. A final simplification rule at the end can be
used to remove the projection constraint from the store.

The following example shows how to project away all \texttt{leq/2} constraints
that involve arguments not contained in a given set \texttt{Vars}:
\begin{Verbatim}
project(Vars) \ leq(X,Y) <=> 
                \+ (member(X,Vars),member(Y,Vars)) | true.
project(Vars)            <=> true.
\end{Verbatim}
Besides removal of constraints more sophisticated operations such as weakening
are possible. E.g. consider a set solver with two constraints:
\texttt{in/2} that requires an element to be in a set and \texttt{nonempty/1}
that requires a set to be non-empty. The rules for projection could
include the following weakening rule:
\begin{Verbatim}
project(Vars) \ in(Elem,Set) <=> 
                member(Set,Vars), 
                \+ member(Elem,Vars) | nonempty(Set).
\end{Verbatim}

\section{Answer Set Optimization}\label{sec:asopt}

In this section we consider various ways for reducing the size of the
answer set. First, in Section \ref{sec:asopt:subsumption} we consider the
subsumption-based technique proposed by Toman. This leads us to a sidetrack
in Section \ref{sec:asopt:dynprog} where we outline a technique for dynamic
programming through answer subsumption. Section \ref{sec:asopt:general}
continues with the main story: it is established that answer subsumption
is suboptimal for general constraint domains and a generalized approach is
proposed instead. In Section \ref{sec:asopt:relaxed} we relax the soundness
condition of answer reduction and speculate on applications to program
analysis. Finally, in Section \ref{sec:asopt:evaluation}, we evaluate the
two main approaches.

\subsection{Answer Subsumption}\label{sec:asopt:subsumption}

Some of the answers computed for a tabled predicate may be
redundant and so need not be saved. The property is exploited by
Definition~\ref{def:tabling:slgc:entailment}, the Optimized Answer Set
definition.  In terms of \slg{\Herbrand}, consider for example that the answer
\texttt{p(a,X)} is already in the table of predicate \texttt{p/2}. Now a new
answer, \texttt{p(a,b)} is found. This new answer is redundant as it is covered
by the more general \texttt{p(a,X)} that is already in the table. Hence it is
logically valid to not record this answer in the table, and to simply discard
it. This does not affect the soundness or completeness of the procedure.


We can extend the idea of answer subsumption to CHR
constraints. This path length computation will serve as an illustration:
\begin{example}
\begin{verbatim}
     dist(A,B,D) :- edge(A,B,D1), leq(D1,D).
     dist(A,B,D) :- dist(A,C,D1), edge(C,B,D2), leq(D1 + D2, D).
\end{verbatim}
Suppose appropriate rules for the \texttt{leq/2} constraint are in the above
program, where \texttt{leq} means less-than-or-equal. The semantics are that
\texttt{dist(A,B,D)} holds if there is a path from \texttt{A} to \texttt{B}
of length less than or equal to \texttt{D}. In other words, \texttt{D}
is an upper bound on the length of a path from \texttt{A} to \texttt{B}.

If the answer \texttt{dist(n1,n2,D) :- leq(d1, D)} is already in the table
and a new answer \texttt{dist(n1,n2,D) :- leq(d2, D)}, where \texttt{d1 =<
d2}, is found, then this new answer is redundant. Hence it can be discarded.
This does not affect the soundness, since logically the same answers
are covered.
\end{example}

A strategy for establishing implication is provided by the following property:
\begin{equation}\label{prop:implication}
\forall i \in \{0,1\}: C_{1-i} \rightarrow C_i \Longleftrightarrow C_0 \wedge C_1 \leftrightarrow C_i
\end{equation}
for any logical formulas $C_0$ and $C_1$.
In particular, consider  $C_0$ and $C_1$ to be a previous answer constraint
store and a newly computed one.
The strategy then is as follows. At the end of the tabled predicate's execution a previous
answer store $C_0$ is merged with a new answer store $C_1$. After merging, the
store is simplified and propagated to $C$ by the available rules of the CHR
program \Prog. This combines the two answers into a new one. This mechanism can
be used to check entailment of one answer by the other: if the combined answer
store $S$ is equal to one of the two, then that answer store entails the other.

A practical procedure is the following:
\begin{Verbatim}
        :- table tabled_p/3.

        tabled_p(X,Y,NStoreEncoding) :-
          original_p(X,Y),
          project([X,Y]),
          encode_store(StoreEncoding),
          ( previous_answer(p(X,Y,PrevStoreEncoding),AnswerID),
            decode_store(PrevStoreEncoding),
            encode_store(Conjunction),
            ( Conjunction == PrevStoreEncoding ->
                del_answer(AnswerID),
                fail
            ;
                Conjunction \== StoreEncoding
            ) ->
                fail 
          ;
                NStoreEncoding = StoreEncoding
          ),
          empty_store.
\end{Verbatim}
After computing, projecting and Herbrand encoding a new answer store $C_1$,
we look at previous answer stores $C_0$. We assume that there is a built-in
predicate \texttt{previous\_answer/3} for this purpose, that backtracks
over previous answers and also provides a handle \texttt{AnswerID} to the
returned answer. As the previous answer store is still in Herbrand encoding, we
decode it.  This has the simultaneous effect of adding it to the new implicit
CHR constraint store that is still in place, i.e. it computes $C_0 \wedge
C_1$. This resulting conjunction is Herbrand encoded for further comparison.
Syntactical equality ($\equiv$) is used as a sound approximation of the
equivalence check for the first equivalence sign ($\leftrightarrow$)
in the Formula \ref{prop:implication}. If the conjunction equals the
previous answer \texttt{PrevStoreEncoding}, then that previous answer is
implied by the new answer and hence obsolete. We use the built-in predicate
\texttt{del\_answer} to erase it from the answer table and we backtrack
over alternative previous answers. Otherwise, if the conjunction does not
equal the new answer, then neither implies the other and we also backtrack
over alternative previous answers. However, if the conjunction equals the
new answer \texttt{StoreEncoding}, that means it is implied by the previous
answer. Hence we fail, ignoring further alternative previous answers. If on
the other hand, the resulting answer is not implied by any previous answers,
then it is a genuinely new answer and is stored in the answer table.

\begin{example}
Consider again the \texttt{dist/3} example, and assume that the answer stores
\texttt{\{leq(7,D\}}, \texttt{\{leq(3,D))\}} and \texttt{\{leq(5,D))\}} are
successively produced for the query \texttt{?- dist(a,b,D)}. When the first answer,
\texttt{\{leq(7,D\}} is produced, there are no previous answers, so it makes its way
into the answer table. For the second answer, \texttt{\{leq(3,D))\}} there is already 
a previous answer \texttt{\{leq(7,D\}}, so both are conjoined.
The following rule \texttt{leq/2} rule simplifies the conjunction
to retain the more general answer:
\begin{center}
\begin{BVerbatim}
leq(N1,D) \ leq(N2,D) <=> N1 >= N2 | true.
\end{BVerbatim}
\end{center}
Hence, the resulting solved form of the conjunction is \texttt{\{leq(D,7)\}}, or in other words
the previous answer. In other words, this previous answer is implied by the new answer. So it is deleted from 
the answer table and the new answer is recorded. Finally, following the same procedure we
discover that the third answer is already implied by the second one. So the final
answer set contains just the second answer.

Note that the \texttt{dist/3} program would normally generate an
infinite number of answers for a cyclic graph, logically correct but not
terminating. However, if it is tabled with answer subsumption, it does
terminate for non-negative weights. Not only does it terminate, it only
produces one answer, namely \texttt{dist(n1,n2,D) :- leq(d,D)} with \texttt{d}
the length of the shortest path. Indeed, the predicate only returns the
optimal answer.
\end{example}

The syntactical equality check on the Herbrand encoding is in general only an
approximation of a proper equivalence check. An option for the \texttt{table\_chr} declarations allows to improve
its effectiveness: \texttt{canonical\_form(\emph{PredName})} specifies the
name of the predicate that should compute the (approximate) canonical form
of the Herbrand encoded answer constraint store. This canonical form is used
to check equivalence of two constraint stores.
\begin{example}
Both \texttt{[leq(1,X),leq(X,3)]} and \texttt{[leq(X,3),leq(1,X)]} are permutations
of the same Herbrand constraint store encoding. Obviously, based on a simple syntactic
equality check they are different. However, they can both be reduced to the same canonical form,
e.g. with the help of the Prolog built-in \texttt{sort/2}.
\end{example}

We refer to \cite{tom:implication:rule2005} for a more elaborated discussion
of the property \ref{prop:implication} and an alternative, more elaborate
implementation of the implication checking strategy in CHR.

In contrast to our generic approach above, the traditional approach in CLP
is for the solver to provide a number of predefined {\em ask} constraints \cite{saraswat:ccp},
i.e. subsumption checks for primitive constraints. These primitive ask
constraints can then be combined to form more complicated subsumption checks \cite{duck:ask}.
We have avoided this approach because it puts a greater burden on the
constraint solver implementer, who has to provide the implementation of the
primitive ask constraints. In future work, we could incorporate user-defined
ask constraints in our generic approach for greater programmer control over
performance and accuracy of subsumption tests. 

\subsection{Dynamic Programming through Answer Subsumption}\label{sec:asopt:dynprog}

The technique used in the \texttt{dist/3} program is to replace the computation
of the exact distance of a path with the computation of an upper bound on the
distance via constraints. Then, by tabling the predicate and performing answer
subsumption, the defining predicate has effectively been turned into an
optimizing one, computing the length of the shortest path. It is a
straightforward yet powerful optimization technique that can be applied to
other defining predicates as well, turning them into optimizing (dynamic
programming) predicates with a minimum of changes. 

In comparison, the usual approach consists in explicitly computing the list of all
answers, e.g. using Prolog's \texttt{findall/3} meta-programming built-in, and in
processing this list of answers.
Guo and Gupta \cite{gupta:dynamic_programming} have added a specific feature to
tabled execution to realize this dynamic programming functionality. In adding
support for CHR to tabling, we get this functionality for free.

\subsection{General Answer Compaction}\label{sec:asopt:general}

Definition \ref{def:tabling:slgc:entailment} yields a sound approach for
reducing the size of answer tables. However, we have discovered that it 
is only a special case of what is really possible.
Therefore, we propose the following generalized definition of answer sets, {\em compacted
answer set}, which covers all sound approaches for reducing the answer set size. 

%

\begin{definition}[Compacted Answer Set]
A compacted answer set of the query $(G,C,P)$, denoted $\ans(G,C)$, is a set such that:
\begin{itemize}
\item No new fully instantiated (i.e. ground) answers are introduced:
\begin{equation}\label{prop:nonew}
\begin{array}{c}
\forall A, \theta: (A \in \ans(G,C)) \wedge (\CT \vdash A\theta) \\ 
\Longrightarrow \\
\exists A',\theta': (\texttt{ans}(G;A') \in \mathit{slg}(G,C)) \wedge (A'\theta' \equiv A\theta)
\end{array}
\end{equation}
\item All fully instantiated answers are covered:
\begin{equation}\label{prop:coverage}
\begin{array}{c}
\forall A', \theta': (\texttt{ans}(G;A') \in \mathit{slg}(G,C)) \wedge (\CT \vdash A'\theta) \\ 
\Longrightarrow \\
\exists A, \theta: (A \in \ans(G,C)) \wedge (A\theta \equiv A'\theta')
\end{array}
\end{equation}
\item The answer set is more compact than the individual answers:
\begin{equation}
\#\ans(G,C) \leq \#\{ A' | \texttt{ans}(G;A') \in \mathit{slg}(G,C) \}
\end{equation}
\end{itemize}
where $A$ and $A'$ are constraint stores and $\theta$ and $\theta'$ are valuations.
\end{definition}

Note that an optimized answer set is a special instance of a compacted
answer set and certainly, for Herbrand constraints, it is an optimal
strategy, because:
\begin{equation}
\Prolog \models \forall H_0, H_1, H: H \leftrightarrow H_0 \vee H_1 \Longrightarrow \exists i \in \{0,1\}: H \leftrightarrow H_i
\end{equation}
where $H_0,H_1,H$ are conjunctions of Herbrand equality constraints.
In other words, for finding a single Herbrand constraint that covers two
given ones, it is sufficient to considering those two.

Unfortunately, a similar property does not hold for all constraint domains: a single constraint
store may be equivalent to the disjunction of two others, while it is not
equivalent to either of the two.  For example, $\mathit{leq(X,Y) \vee leq(Y,X)
\leftrightarrow true}$ and yet we have that neither $\mathit{leq(X,Y)
\leftrightarrow true}$ nor $\mathit{leq(Y,X) \leftrightarrow true}$.

Nevertheless, checking whether one answer subsumes the other is a rather
convenient strategy, since it does not require any knowledge on the
particularities of the used constraint solver. That makes it a good choice
for the default strategy for CHR answer subsumption. Better strategies
may be supplied for particular constraint solvers through the option
\texttt{answer\_combination(\emph{PredName})}. It specifies the name of the
predicate that returns the disjunction of two given answer stores, or fails
if it cannot find one. 

\begin{example}\label{ex:union}
Consider a simple interval-based solver, featuring constraints of the form $X \in [L,U]$,
where $X$ is a constraint variable and $L$ and $U$ are integers,
and the rules:
\begin{Verbatim}[commandchars=\\\{\},codes={\catcode`$=3}]
        X $\in$ [L,U] ==> L =< U.
        X $\in$ [L1,U1], X $\in$ [L2,U2] <=> X $\in$ ([L1,U1] $\cap$ [L2,U2]).
\end{Verbatim}

For this solver, the subsumption approach merges two constraints $X \in [L_1,U_1]$
and $X \in [L_2,U_2]$ iff $[L_1,U_1] \subseteq [L_2,U_2]$ or $[L_2,U_2] \subseteq [L_1,U_1]$.
However, it fails to work for e.g. $X \in [1,3]$ and $X \in [2,4]$. Nevertheless there is a
single constraint form that covers both: $X \in [1,4]$. An optimal answer combinator in this case
is one that returns the union of two overlapping intervals. This also captures the subsumption approach.
If the intervals do not overlap, there is no single constraint that covers
both without introducing new answers.
\end{example}

Note that the idea of general answer compaction is not specific implementation
of constraints, and, in particular, should apply to non-CHR constraint solvers too.

%
%


\subsection{Relaxed Answer Compaction Semantics}\label{sec:asopt:relaxed} 

For some applications the soundness condition of answer generalization
can be relaxed. An example in regular
Prolog would be to have two answers \texttt{p(a,b)} and \texttt{p(a,c)} and
to replace the two of them with one answer \texttt{p(a,X)}. This guarantees
(for positive programs) that no answers are lost, but it may introduce
extraneous answers. In other words, property \ref{prop:coverage} is preserved
while property \ref{prop:nonew} is not.
A similar technique is possible with constrained answers.
While this approach is logically unsound, it may be acceptable for some
applications if only answer coverage is required.

An example is the use of the least upper bound (lub) operator to
combine answers in the tabled abstract interpretation setting of
\cite{bart:abstract_interpretation}. There is often a trade-off between
accuracy and efficiency in space and time. By exploiting this trade-off
abstract interpretation can remain feasible in many circumstances.

Toman has explored in \cite{toman:analysis} the use of CLP for program analysis
and compared it to abstract interpretation. In his proposal constraints serve
as the abstractions of concrete values, and bottom-up computation or tabling
is necessary to reach a fixpoint over recursive program constructs. He notes
that the CLP approach is less flexible than actual abstract interpretation
because it lacks flexible control over the accuracy/efficiency trade-off.
We believe that our proposal for relaxed answer compaction could function
as a lub or widening operator to remedy this issue, making Toman's program
analysis technique more practical. This remains to be explored in future work.

\subsection{Evaluation: A Shipment Problem}\label{sec:asopt:evaluation}

We evaluate the usefulness of the two proposed answer set optimization approaches
based on a shipment problem.

\textbf{Problem statement:} \emph{
There are N packages available for shipping using trucks. Each package has
a weight and some constraints on the time to be delivered. Each truck has
a maximum load and a destination. Determine whether there is a subset of
the packages that can fully load a truck destined for a certain place
so that all the packages in this subset are delivered on time.
} (from \cite{bao:thesis})

The problem is solved by the \emph{truckload} program:

\begin{Verbatim}[frame=single,framesep=1mm,label=The \texttt{truckload} Program]
:- constraints leq/2.

leq(X,X)              <=> true.
leq(N1,N2)            <=> number(N1), number(N2) | N1 =< N2.
leq(N1,X) \ leq(N2,X) <=> number(N1), number(N2), N1 > N2 | true.
leq(X,N1) \ leq(X,N2) <=> number(N1), number(N2), N1 < N2 | true.
leq(X,Y)  \ leq(X,Y)  <=> true.
leq(X,Y)  , leq(Y,Z)  ==> leq(X,Z).

truckload(0,0,_,_).
truckload(I,W,D,T) :-                   % do not include pack I
	I > 0,
	I1 is I - 1,
	truckload(I1,W,D,T).
truckload(I,W,D,T) :-                   % include pack I
	I > 0,
	pack(I,Wi,D,T),
	W1 is W - Wi,
	W1 >= 0,
	I1 is I - 1,
	truckload(I1,W1,D,T).

pack(30,29,chicago,T) :- leq(19,T),leq(T,29).
pack(29,82,chicago,T) :- leq(20,T),leq(T,29).
pack(28,24,chicago,T) :- leq(8,T),leq(T,12).
%...
pack(3,60,chicago,T) :- leq(4,T),leq(T,29).
pack(2,82,chicago,T) :- leq(28,T),leq(T,29).
pack(1,41,chicago,T) :- leq(27,T),leq(T,28).
\end{Verbatim}

Packages are represented by a {\em constraint database}: clauses of \texttt{pack/4}, e.g.
\begin{center}
\begin{BVerbatim}
pack(3,60,chicago,T) :- leq(4,T),leq(T,29).
\end{BVerbatim}
\end{center}
means that the third package weights 60 pounds, is destined for Chicago
and has to be delivered between the 4th and the 29th day.
The \texttt{truckload/4} predicate computes the answer to the problem, e.g.
\texttt{:- truckload(30,100,chicago,T)} computes whether a subset of the packages
numbered 1 to 30 exists to fill up a truck with a maximum load of 100 pounds
destined for Chicago. The time constraints are captured in the bound on the
constraint variable \texttt{T}. There may be multiple answers to this query,
if multiple subsets exist that satisfy it.

We have run the program in four different modes:
\begin{itemize}
\item {\sc No Tabling}: the program is run as is without tabling.
\item {\sc Tabling - Plain}: to avoid the recomputation of subproblems in recursive calls
      the \texttt{truckload/4} predicate is tabled with:
\begin{verbatim}
    :- table_chr truckload(_,_,_,chr) 
                 with [encoding(goal)].
\end{verbatim}
\item {\sc Tabling - Sorted}: the answer store is canonicalized by simple sorting such that
      permutations are detected to be identical answers:
\begin{verbatim}
    :- table_chr truckload(_,_,_,chr) 
                 with [encoding(goal),
                       canonical_form(sort)].
\end{verbatim}
\item {\sc Tabling - Combinator}: we apply the custom answer combinator proposed in Example \ref{ex:union}: two answers with
      overlapping time intervals are merged into one answer with the union of the time intervals. 
      This variant is declared as:
\begin{verbatim}
    :- table_chr truckload(_,_,_,chr) 
                 with [encoding(goal),
                       answer_combination(interval_union)].
\end{verbatim}
with \texttt{interval\_union/3} the custom answer combinator.
\end{itemize}

\begin{table}
\begin{center}
\begin{tabular}{|l||r|r|r|r|}
\cline{1-5}
        & {\sc No Tabling} & \multicolumn{3}{c|}{{\sc Tabling}} \\
Load    &            & {\sc Plain}  & {\sc Sorted}        & {\sc Combinator} \\
\cline{1-5}
100 &         $<$1     &   100 &   100 &   100 \\
200 &        160     &   461 &   461 &   451 \\
300 &      2,461     & 1,039 & 1,041 &   971 \\
400 &     12,400     & 1,500 & 1,510 & 1,351 \\
500 & $>$ 5 min.     & 1,541 & 1,541 & 1,451 \\
\cline{1-5}
\end{tabular}
\end{center}
\caption{\label{tab:truckload:runtime} Runtime results for the \texttt{truckload} program}
\end{table}
\begin{table}
\begin{center}
\begin{tabular}{|l||r|r|r|}
\cline{1-4}
        & \multicolumn{3}{c|}{{\sc Tabling}} \\
Load    & {\sc Plain}  & {\sc Sorted}        & {\sc Combinator} \\
\cline{1-4}
100 &   286 &   286 &   279 \\
200 &   979 &   956 &   904 \\
300 & 1,799 & 1,723 & 1,584 \\
400 & 2,308 & 2,202 & 2,054 \\
500 & 2,449 & 2,365 & 2,267 \\
\cline{1-4}
\end{tabular}
\end{center}
\caption{\label{tab:truckload:space} Space usage for the \texttt{truckload} program}
\end{table}
\begin{table}
\begin{center}
\begin{tabular}{|l||r|r|r|}
\cline{1-4}
        & \multicolumn{3}{c|}{tabling} \\
load    & plain  & sorted        & combinator \\
\cline{1-4}
\cline{1-4}
100 &   324 &   324 &   283\\
200 & 2,082 & 2,069 & 1,686\\
300 & 4,721 & 4,665 & 3,543\\
400 & 5,801 & 5,751 & 4,449\\
500 & 4,972 & 4,935 & 4,017\\
\cline{1-4}
\end{tabular}
\end{center}
\caption{\label{tab:truckload:answers} Number of tabled answers for the \texttt{truckload} program}
\end{table}

Table \ref{tab:truckload:runtime} contains the runtime results of running
the program in the four different modes for different maximum loads. Runtime
is in milliseconds and has been obtained on an Intel Pentium 4 2.00 GHz with 512 MB of
RAM, with XSB 6.1 running on Linux 2.6.18. For the modes with tabling the space usage, in
kilobytes, of the tables and number of unique answers have been recorded as
well, in Table \ref{tab:truckload:space} and Table \ref{tab:truckload:answers}
respectively.

It is clear from the results that tabling has an overhead for small
loads, but that it scales much better. Both the modes with the canonical
form and the answer combination have a slight space advantage over plain
tabling which increases with the total number of answers. There is hardly
any runtime effect for the canonical form, whereas the answer combination
mode is faster with increasing load.

In summary, 
canonicalization of the answer store
and answer combination can have a favorable impact on both runtime and
table space depending on the particular problem.

\section{Related and Future Work}\label{sec:related}

The theoretical background for this paper, SLG$^\CT$ resolution, was realized
by Toman in \cite{toman:memo}. Toman establishes soundness, completeness
and termination properties for particular classes of constraint domains.
While he has implemented a prototype implementation of SLG$^\CT$ resolution
for evaluation, no practical and fully-fledged implementation in a Prolog
system was done.

Various ad hoc approaches to using constraints in XSB were used in the past by
Ramakrishnan et al., such as a meta-interpreter \cite{mcrtpre}, interfacing
with a solver written in C \cite{mcrt} and explicit constraint store management
in Prolog \cite{giri}.  However, these approaches are quite cumbersome and lack
the ease of use and generality of CHR.

The most closely related implementation work that this paper builds on is
\cite{bao:tclp}, which presents a framework for constraint solvers written
with attributed variables. Attributed variables are a much cruder tool for
writing constraint solvers though. Implementation issues such as constraint
store encoding and scheduling strategies that are hidden by CHR become
the user's responsibility when she programs with attributed variables. Also
in the tabled setting, the user has to think through all the integration
issues of the attributed variables solver. For CHR we have provided generic
solutions that work for all CHR constraint solvers and more powerful features
can be accessed through parametrized options.

Guo and Gupta propose a technique for dynamic programming with tabling
\cite{gupta:dynamic_programming} that is somewhat similar to the one proposed here.
During entailment checking  a particular argument
in a new answer is compared with the value in the previous answer. Either one is kept
depending on the optimization criterion. Their
technique is specified for particular numeric arguments whereas ours is for
constraint stores and as such more general. Further investigation of our technique is certainly
necessary to establish the extent of its applicability.

Part of this work was previously published at the International Conference of
Logic Programming \cite{tom:xsb:iclp2004} and the Colloquium on  Implementation
of Constraint and Logic Programming Systems \cite{tom:xsb:ciclops2003}.
In \cite{tom:xsb:ciclops2003} we briefly discuss two applications of CHR with tabling
in the field of model checking. The integration of CHR and XSB has shown to
make the implementation of model checking applications with constraints significantly
easier. The next step in the search for applications is to explore
more expressive models to be checked than are currently viable with traditional approaches.

Further applications should also serve to improve the currently limited
performance assessment of CHR with tabling. The shipment problem has given us
some indication of improved performance behavior in practice, but theoretical
reasoning indicates that slow-downs are a possibility as well.

The global CHR store has proven to be one of the main complications in tabling
CHR constraints. For particular CHR programs it is possible to replace the
global data structure with localized, distributed ones. Assessment
\cite{beata} of this approach has shown to be very promising.

Partial abstraction and subsumption are closely related. The former transforms
a call into a more general call while the latter looks for answers to more
general calls, but if none are available still executes the actual call.  We
still have to look at how to implement partial abstraction and the implications
of variant and subsumption based tabling \cite{xsb:subsumption}. 

Finally, better automatic techniques for entailment testing, such as those of
\cite{tom:implication:rule2005}, and for projection should be investigated
in the context of SLG$^\CT$.

\section{Conclusion}\label{sec:conclusion}

We have presented a high-level framework for tabled CLP, based on
a light-weight integration of CHR with a tabled LP system.
Tabling-related problems that have to be solved time and again for
ad hoc constraint solver integrations are solved once and for all for CHR
constraint solvers. 
Solutions have been formulated for call abstraction,
tabling constraint stores, answer projection, answer combination (e.g. for
optimization), and answer set optimization. Hence integrating a particular
CHR constraint solver requires much less knowledge of implementation
intricacies and decisions can be made on a higher level.

If performance turns out to be a bottleneck, once the high-level integration
is stable and well-understood, then its implementation may be specialized in
a lower-level language using the TCHR implementation as its specification.
Our novel contribution, generalized answer set compaction, may certainly 
contribute towards that end.

Finally, we would like to mention that an XSB release, number 2.7, with the
presented CHR system integrated with tabling is publicly available since
December 30, 2004 (see \texttt{http://xsb.sf.net}).

\section*{Acknowledgements}

We are grateful to Beata Sarna-Starosta, Giridhar Pemmasani and
C.R. Ramakrishnan for the interesting discussions and help on applications
of tabled execution and constraints in the field of model checking.

We thank the anonymous reviewers for their helpful comments.


\bibliographystyle{acmtrans} 
\bibliography{tplp}

\begin{thebibliography}{}

\bibitem[\protect\citeauthoryear{Clark}{Clark}{1987}]{clark:completion}
{\sc Clark, K.~L.} 1987.
\newblock {Negation as Failure}.
\newblock In {\em Logic and Databases}, {H.~Gallaire} {and} {J.~Minker}, Eds.
  Plenum Press, New York, 293--322.

\bibitem[\protect\citeauthoryear{Codish, Demoen, and Sagonas}{Codish
  et~al\mbox{.}}{1998}]{bart:abstract_interpretation}
{\sc Codish, M.}, {\sc Demoen, B.}, {\sc and} {\sc Sagonas, K.} 1998.
\newblock {Semantic-based Program Analysis for Logic-based Languages Using
  XSB}.
\newblock {\em International Journal of Software Tools for Technology
  Transfer\/}~{\em 2,\/}~1 (Jan.), 29--45.

\bibitem[\protect\citeauthoryear{Cui}{Cui}{2000}]{bao:thesis}
{\sc Cui, B.} 2000.
\newblock {A System for Tabled Constraint Logic Programming}.
\newblock Ph.D. thesis, State University of New York at Stony Brook.

\bibitem[\protect\citeauthoryear{Cui and Warren}{Cui and
  Warren}{2000a}]{bao:tclp}
{\sc Cui, B.} {\sc and} {\sc Warren, D.~S.} 2000a.
\newblock {A System for Tabled Constraint Logic Programming}.
\newblock In {\em CL 2000: Proceedings of the 1st International Conference on
  Computational Logic}, {J.~W. Lloyd}, {V.~Dahl}, {U.~Furbach}, {M.~Kerber},
  {K.-K. Lau}, {C.~Palamidessi}, {L.~M. Pereira}, {Y.~Sagiv}, {and} {P.~J.
  Stuckey}, Eds. Lecture Notes in Computer Science, vol. 1861. Springer Verlag,
  London, UK, 478--492.

\bibitem[\protect\citeauthoryear{Cui and Warren}{Cui and
  Warren}{2000b}]{bao:attvars}
{\sc Cui, B.} {\sc and} {\sc Warren, D.~S.} 2000b.
\newblock {Attributed Variables in XSB}.
\newblock In {\em Electronic Notes in Theoretical Computer Science}, {I.~Dutra}
  {et~al\mbox{.}}, Eds. Vol.~30. Elsevier, 67--80.

\bibitem[\protect\citeauthoryear{Demoen}{Demoen}{2004}]{hprolog}
{\sc Demoen, B.} 2004.
\newblock {hProlog}.
\newblock http://www.cs.kuleuven.be/\~{}bmd/hProlog/.

\bibitem[\protect\citeauthoryear{Du, Ramakrishnan, and Smolka}{Du
  et~al\mbox{.}}{2000}]{mcrt}
{\sc Du, X.}, {\sc Ramakrishnan, C.~R.}, {\sc and} {\sc Smolka, S.~A.} 2000.
\newblock {Tabled Resolution + Constraints: A Recipe for Model Checking
  Real-Time Systems}.
\newblock In {\em IEEE Real Time Systems Symposium}. Orlando, Florida,
  175--184.

\bibitem[\protect\citeauthoryear{Duck, {Garc\'{\i}a de la Banda}, and
  Stuckey}{Duck et~al\mbox{.}}{2004}]{duck:ask}
{\sc Duck, G.~J.}, {\sc {Garc\'{\i}a de la Banda}, M.}, {\sc and} {\sc Stuckey,
  P.~J.} 2004.
\newblock {Compiling Ask Constraints}.
\newblock In {\em ICLP'04: {P}roceedings of the 20th {I}nternational
  {C}onference on {L}ogic {P}rogramming}. Lecture Notes in Computer Science,
  vol. 3132. Springer Verlag, St-Malo, France, 105--119.

\bibitem[\protect\citeauthoryear{Duck, Stuckey, {Garc\'{\i}a de la Banda}, and
  Holzbaur}{Duck et~al\mbox{.}}{2004}]{duck:refined}
{\sc Duck, G.~J.}, {\sc Stuckey, P.~J.}, {\sc {Garc\'{\i}a de la Banda}, M.},
  {\sc and} {\sc Holzbaur, C.} 2004.
\newblock {The Refined Operational Semantics of Constraint Handling Rules}.
\newblock In {\em ICLP'04: {P}roceedings of the 20th {I}nternational
  {C}onference on {L}ogic {P}rogramming}. Lecture Notes in Computer Science,
  vol. 3132. Springer Verlag, St-Malo, France, 90--104.

\bibitem[\protect\citeauthoryear{Fr{\"u}hwirth}{Fr{\"u}hwirth}{1998}]{thom:mai%
n}
{\sc Fr{\"u}hwirth, T.} 1998.
\newblock Theory and practice of constraint handling rules.
\newblock {\em Journal of {Logic} {Programming}\/}~{\em 37,\/}~1--3 (October),
  95--138.

\bibitem[\protect\citeauthoryear{Fr{\"u}hwirth and Abdennadher}{Fr{\"u}hwirth
  and Abdennadher}{2003}]{thom:book}
{\sc Fr{\"u}hwirth, T.} {\sc and} {\sc Abdennadher, S.} 2003.
\newblock {\em {Essentials of Constraint Programming}}.
\newblock Cognitive Technologies. Springer Verlag.

\bibitem[\protect\citeauthoryear{Guo and Gupta}{Guo and
  Gupta}{2004}]{gupta:dynamic_programming}
{\sc Guo, H.-F.} {\sc and} {\sc Gupta, G.} 2004.
\newblock {Simplifying Dynamic Programming via Tabling}.
\newblock In {\em Proc. Sixth International Symposium on Practical Aspects of
  Declarative Languages}, {P.~V. Hentenryck}, Ed. Lecture Notes in Computer
  Science, vol. 3819. Springer Verlag, 163--177.

\bibitem[\protect\citeauthoryear{Holzbaur}{Holzbaur}{1992}]{christian:attvars}
{\sc Holzbaur, C.} 1992.
\newblock {Metastructures vs. Attributed Variables in the Context of Extensible
  Unification}.
\newblock Tech. Rep. TR-92-23, Austrian Research Institute for Artificial
  Intelligence, Vienna, Austria.

\bibitem[\protect\citeauthoryear{Holzbaur and Fr{\"u}hwirth}{Holzbaur and
  Fr{\"u}hwirth}{2000}]{christian:system}
{\sc Holzbaur, C.} {\sc and} {\sc Fr{\"u}hwirth, T.} 2000.
\newblock {A Prolog Constraint Handling Rules Compiler and Runtime System}.
\newblock {\em Special Issue Journal of Applied Artificial Intelligence on
  Constraint Handling Rules\/}~{\em 14,\/}~4 (April), 369--388.

\bibitem[\protect\citeauthoryear{Jaffar and Lassez}{Jaffar and
  Lassez}{1987}]{jaffar:fixpoint}
{\sc Jaffar, J.} {\sc and} {\sc Lassez, J.-L.} 1987.
\newblock {Constraint Logic Programming}.
\newblock In {\em POPL '87: Proceedings of the 14th ACM SIGACT-SIGPLAN
  symposium on Principles of programming languages}. ACM Press, New York, NY,
  USA, 111--119.

\bibitem[\protect\citeauthoryear{Jaffar and Maher}{Jaffar and
  Maher}{1994}]{jaffar:clp}
{\sc Jaffar, J.} {\sc and} {\sc Maher, M.~J.} 1994.
\newblock {Constraint Logic Programming: A Survey}.
\newblock {\em Journal of Logic Programming\/}~{\em 19/20}, 503--581.

\bibitem[\protect\citeauthoryear{Kanellakis, Kuper, and Revesz}{Kanellakis
  et~al\mbox{.}}{1995}]{kanellakis:cql}
{\sc Kanellakis, P.~C.}, {\sc Kuper, G.~M.}, {\sc and} {\sc Revesz, P.~Z.}
  1995.
\newblock {Constraint query languages}.
\newblock In {\em Selected papers of the 9th annual ACM SIGACT-SIGMOD-SIGART
  symposium on Principles of database systems}. Academic Press, Inc., Orlando,
  FL, USA, 26--52.

\bibitem[\protect\citeauthoryear{Marriott and Stuckey}{Marriott and
  Stuckey}{1998}]{stuckey:clp}
{\sc Marriott, K.} {\sc and} {\sc Stuckey, P.~J.} 1998.
\newblock {\em {Programming with Constraints: an Introduction}}.
\newblock {MIT Press}.

\bibitem[\protect\citeauthoryear{Mukund, Ramakrishnan, Ramakrishnan, and
  Verma}{Mukund et~al\mbox{.}}{2000}]{mcrtpre}
{\sc Mukund, M.}, {\sc Ramakrishnan, C.~R.}, {\sc Ramakrishnan, I.~V.}, {\sc
  and} {\sc Verma, R.} 2000.
\newblock {Symbolic Bisimulation using Tabled Constraint Logic Programming}.
\newblock In {\em International Workshop on Tabulation in Parsing and
  Deduction}. Vigo, Spain, 1--9.

\bibitem[\protect\citeauthoryear{Pemmasani, Ramakrishnan, and
  Ramakrishnan}{Pemmasani et~al\mbox{.}}{2002}]{giri}
{\sc Pemmasani, G.}, {\sc Ramakrishnan, C.~R.}, {\sc and} {\sc Ramakrishnan,
  I.~V.} 2002.
\newblock {Efficient Model Checking of Real Time Systems Using Tabled Logic
  Programming and Constraints}.
\newblock In {\em International Conference on Logic Programming}. Lecture Notes
  in Computer Science. Springer, Copenhagen, Denmark, 405--410.

\bibitem[\protect\citeauthoryear{Rao, Ramakrishnan, and Ramakrishnan}{Rao
  et~al\mbox{.}}{1996}]{xsb:subsumption}
{\sc Rao, P.}, {\sc Ramakrishnan, C.~R.}, {\sc and} {\sc Ramakrishnan, I.~V.}
  1996.
\newblock {A Thread in Time Saves Tabling Time}.
\newblock In {\em Joint International Conference and Symposium on Logic
  Programming}. 112--126.

\bibitem[\protect\citeauthoryear{Saraswat and Rinard}{Saraswat and
  Rinard}{1990}]{saraswat:ccp}
{\sc Saraswat, V.~A.} {\sc and} {\sc Rinard, M.} 1990.
\newblock Concurrent constraint programming.
\newblock In {\em POPL '90: Proceedings of the 17th ACM SIGPLAN-SIGACT
  symposium on Principles of programming languages}. ACM Press, New York, NY,
  USA, 232--245.

\bibitem[\protect\citeauthoryear{Sarna-Starosta and
  Ramakrishnan}{Sarna-Starosta and Ramakrishnan}{2003}]{beata:efa}
{\sc Sarna-Starosta, B.} {\sc and} {\sc Ramakrishnan, C.~R.} 2003.
\newblock {Constraint-Based Model Checking of Data-Independent Systems}.
\newblock In {\em 5th International Conference on Formal Engineering Methods,
  ICFEM 2003}, {J.~S. Dong} {and} {J.~Woodcock}, Eds. Lecture Notes in Computer
  Science, vol. 2885. Springer-Verlag, 579--598.

\bibitem[\protect\citeauthoryear{Sarna-Starosta and
  Ramakrishnan}{Sarna-Starosta and Ramakrishnan}{2007}]{beata}
{\sc Sarna-Starosta, B.} {\sc and} {\sc Ramakrishnan, C.~R.} 2007.
\newblock {Compiling Constraint Handling Rules for Efficient Tabled
  Evaluation}.
\newblock In {\em PADL'07: Ninth International Symposium on Practical Aspects
  of Declarative Languages}, {M.~Hanus}, Ed. Lecture Notes in Computer Science.
  Springer Verlag, 170--184.

\bibitem[\protect\citeauthoryear{Schrijvers}{Schrijvers}{2005}]{tom:phd}
{\sc Schrijvers, T.} 2005.
\newblock {Analyses, Optimizations and Extensions of Constraint Handling
  Rules}.
\newblock Ph.D. thesis, Department of Computer Science, K.U.Leuven, Leuven,
  Belgium.

\bibitem[\protect\citeauthoryear{Schrijvers and Demoen}{Schrijvers and
  Demoen}{2004}]{tom:kulchr}
{\sc Schrijvers, T.} {\sc and} {\sc Demoen, B.} 2004.
\newblock {T}he {K}.{U}.{L}euven {CHR} system: {I}mplementation and
  application.
\newblock In {\em {First Workshop on Constraint Handling Rules: Selected
  Contributions}}, {T.~Fr{\"{u}}hwirth} {and} {M.~Meister}, Eds. Ulm, Germany,
  1--5.

\bibitem[\protect\citeauthoryear{Schrijvers, Demoen, Duck, Stuckey, and
  Fr{\"{u}}hwirth}{Schrijvers et~al\mbox{.}}{2006}]{tom:implication:rule2005}
{\sc Schrijvers, T.}, {\sc Demoen, B.}, {\sc Duck, G.}, {\sc Stuckey, P.}, {\sc
  and} {\sc Fr{\"{u}}hwirth, T.} 2006.
\newblock {Automatic Implication Checking for {CHR} Constraints}.
\newblock In {\em Electronic Notes in Theoretical Computer Science}. Vol. 147.
  93--111.

\bibitem[\protect\citeauthoryear{Schrijvers and Fr{\"u}hwirth}{Schrijvers and
  Fr{\"u}hwirth}{2006}]{tom:unionfind:pearl}
{\sc Schrijvers, T.} {\sc and} {\sc Fr{\"u}hwirth, T.} 2006.
\newblock {Optimal Union-Find in Constraint Handling Rules}.
\newblock {\em Theory and Practice of Logic Programming\/}~{\em 6,\/}~1\&2,
  213--224.

\bibitem[\protect\citeauthoryear{Schrijvers and Warren}{Schrijvers and
  Warren}{2004}]{tom:xsb:iclp2004}
{\sc Schrijvers, T.} {\sc and} {\sc Warren, D.~S.} 2004.
\newblock {C}onstraint handling rules and tabled execution.
\newblock In {\em ICLP'04: {P}roceedings of the 20th {I}nternational
  {C}onference on {L}ogic {P}rogramming}, {B.~Demoen} {and} {V.~Lifschitz},
  Eds. Lecture Notes in Computer Science, vol. 3132. Springer Verlag, St-Malo,
  France, 120--136.

\bibitem[\protect\citeauthoryear{Schrijvers, Warren, and Demoen}{Schrijvers
  et~al\mbox{.}}{2003}]{tom:xsb:ciclops2003}
{\sc Schrijvers, T.}, {\sc Warren, D.~S.}, {\sc and} {\sc Demoen, B.} 2003.
\newblock {CHR} for {XSB}.
\newblock In {\em {CICLOPS} 2003: {P}roceedings of the {C}olloquium on
  {I}mplementation of {C}onstraint and {LO}gic {P}rogramming {S}ystems},
  {R.~Lopes} {and} {M.~Ferreira}, Eds. University of Porto, Mumbai, India,
  7--20.

\bibitem[\protect\citeauthoryear{Toman}{Toman}{1996}]{toman:wfs}
{\sc Toman, D.} 1996.
\newblock {Computing the Well-founded Semantics for Constraint Extensions of
  Datalog}.
\newblock In {\em Proceedings of CP'96 Workshop on Constraint Databases}.
  Number 1191 in Lecture Notes in Computer Science. Cambridge, MA, USA, 64--79.

\bibitem[\protect\citeauthoryear{Toman}{Toman}{1997a}]{toman:analysis}
{\sc Toman, D.} 1997a.
\newblock {Constraint Databases and Program Analysis Using Abstract
  Interpretation}.
\newblock In {\em Constraint Databases and Their Applications, Second
  International Workshop on Constraint Database Systems (CDB '97)}, {V.~Gaede},
  {A.~Brodsky}, {O.~G{\"{u}}nther}, {D.~Srivastava}, {V.~Vianu}, {and}
  {M.~Wallace}, Eds. Lecture Notes in Computer Science, vol. 1191. Springer
  Verlag, 246--262.

\bibitem[\protect\citeauthoryear{Toman}{Toman}{1997b}]{toman:memo}
{\sc Toman, D.} 1997b.
\newblock {Memoing Evaluation for Constraint Extensions of Datalog}.
\newblock {\em {Constraints: An International Journal, Special Issue on
  Constraints and Databases}\/}~{\em 2,\/}~3/4 (December), 337--359.

\bibitem[\protect\citeauthoryear{Warren et~al\mbox{.}}{Warren
  et~al\mbox{.}}{2005}]{xsb}
{\sc Warren, D.~S.} {\sc et~al\mbox{.}} 2005.
\newblock {The XSB Programmer's Manual: version 2.7, vols. 1 and 2}.
\newblock http://xsb.sf.net.

\bibitem[\protect\citeauthoryear{Wielemaker}{Wielemaker}{2004}]{swiprolog}
{\sc Wielemaker, J.} 2004.
\newblock {SWI-Prolog release 5.4.0}.
\newblock http://www.swi-prolog.org/.

\bibitem[\protect\citeauthoryear{Wolper}{Wolper}{1986}]{wolper86}
{\sc Wolper, P.} 1986.
\newblock Expressing interesting properties of programs in propositional
  temporal logic.
\newblock In {\em POPL '86: Proceedings of the 13th ACM SIGACT-SIGPLAN
  symposium on Principles of programming languages}. ACM Press, New York, NY,
  USA, 184--193.

\end{thebibliography}

\end{document}